# Giant Purcell broadening and Lamb shift for DNA-assembled near-infrared quantum emitters

Sachin Verlekar,[1,*] Maria Sanz-Paz,[2,*] Mario Zapata-Herrera,[3,4] Mauricio Pilo-Pais,[2] Karol Kołątaj,[2] Ruben Esteban,[3,4,†] Javier Aizpurua,[4,5,6] Guillermo Acuna,[2,‡] and Christophe Galland[1,§]

[1]*Institute of Physics, Ecole Polytechnique Fédérale de Lausanne (EPFL), Switzerland*

[2]*Department of Physics, University of Fribourg, Fribourg CH-1700, Switzerland*

[3]*Materials Physics Center CSIC-UPV/EHU, 20018 Donostia-San Sebastián, Spain*

[4]*Donostia International Physics Center (DIPC), 20018 Donostia-San Sebastián, Spain*

[5]*Ikerbasque, Basque Foundation for Science, 48013 Bilbao, Spain*

[6]*Dept. of Electricity and Electronics, University of the Basque Country (UPV/EHU), 48940 Leioa, Spain*

Controlling the light emitted by individual molecules is instrumental to a number of novel nanotechnologies ranging from super-resolution bio-imaging and molecular sensing to quantum nanophotonics. Molecular emission can be tailored by modifying the local photonic environment, for example by precisely placing a single molecule inside a plasmonic nanocavity with the help of DNA origami. Here, using this scalable approach, we show that commercial fluorophores experience giant Purcell factors and Lamb shifts, reaching values on par with those recently reported in scanning tip experiments. Engineering of plasmonic modes enables cavity-mediated fluorescence far detuned from the zero-phonon-line (ZPL) – at detunings that are up to two orders of magnitude larger than the fluorescence linewidth of the bare emitter and reach into the near-infrared. Our results evidence a regime where the emission linewidth is dominated by the excited state lifetime, as required for indistinguishable photon emission, baring relevance to the development of nanoscale, ultrafast quantum light sources and to the quest toward single-molecule cavity-QED. In the future, this approach may also allow to design efficient quantum emitters at infrared wavelengths, where standard organic sources have a reduced performance.

## INTRODUCTION

Localized surface plasmon resonances (LSPRs) supported by metallic nanoparticles can be used to selectively enhance the scattering and absorption of light, allowing for the realisation of optical nanoantennas [1]. When two nanoparticles, or a nanoparticle and a metal surface, are placed in close proximity, they can support LSPRs whose near field is tightly confined in the separating gap, thus forming plasmonic nanocavities [2–4]. Deep sub-wavelength field confinement in nanocavities allows for the engineering of light-matter interactions at the single molecule level [5–7]; in particular, a quantum emitter placed in a nanocavity experiences a modified spontaneous emission rate, quantified by the local density of optical states (LDOS). In this context, a major enterprise has been to leverage the Purcell effect to increase the radiative decay rates of various quantum emitters coupled to plasmonic nanoantennas and nanocavities [8–13].

For molecular emitters, which show multiple vibronic levels, coupling to LSPRs can significantly reshape their emission spectra when the LSPR is detuned from their zero phonon line (ZPL), even without reaching the strong coupling regime. The effect can be understood as the enhancement of decay rates for vibronic transitions that spectrally overlap with the broad plasmonic resonance. Previous systems employed to study spectral reshaping include lithographically fabricated arrays coated with emitters [14–17], plasmonic antennas immersed in fluorophore solutions [18–22] as well as nanoparticles encapsulated by quantum emitters in a core-shell geometry [23–26]. More recently, spectral reshaping was studied at room temperature on single fluorophores immobilized close to the tip of gold nanorods using DNA-origami [27]. The DNAorigami technique [28] allows for bottom-up fabrication of 3D nanostructures [28–30] with dimensions in the tens to hundreds of nanometers, and has found several applications in nano-photonics [31] as it enables the precise positioning of both metallic nanoparticles and single photon emitters with controlled orientation and stoichiometry [32–34]. Some notable examples include the enhancement of single photon emission [35– 39], surface-enhanced Raman scattering (SERS) [40–42], directing light [43–45], ultra-fast phenomena [46], strong coupling [47, 48] and super-resolution microscopy [49]. In Ref. [27] pronounced reshaping of the light emission was demonstrated when the nanorod LSPR overlaps with the free-space fluorophore spectrum. While most results could be explained by the Purcell effect in the weak coupling regime through Fermi's Golden Rule [23], some discrepancies were noticed [27] and attributed to a possible violation of Kasha's rule:
because of the shortened excited state lifetime, emission might also occur from excited vibrational sub-levels.

All aforementioned studies were performed under ambient conditions so that the natural emission linewidths largely overlapped with the plasmonic mode(s) involved in spectral reshaping, leaving open the question whether emission can be induced well beyond the natural linewidth. Moreover, it is valuable to independently engineer the total vs. the radiative LDOS: while the latter largely determines the reshaping function in the weak coupling limit, the former actually determines the change in excited state lifetime, with its possible impacts on emission linewidth. In theory, a modified LDOS is also accompanied by a modification in the Lamb shift [50] (as they are connected to the imaginary and real part of the dyadic Green's function, respectively), but this effect has remained largely elusive in the context of plasmonic antennas, where it should however be particularly pronounced. Only recently has it been evidenced in a highly controlled environment, with a system consisting of free-base phthalocyanine molecules evaporated on monocrystalline silver films coated with bi- and trilayer NaCl and studied in an ultra-high vacuum, cryogenic scanning tunneling microscope [51, 52]. Such systems contrast with the need for monolithic, solid-state (and ideally mass-producible) cavities that would fuel scalable photonic technologies harvesting plasmonic LDOS control for the engineering of single emitter properties.

Here, we deterministically couple single commercial fluorophores to gold nanodimers (NDs) using DNA origami and study the resulting hybrid structures at cryogenic temperatures, where the ZPL of the bare fluorophore (as deposited on a glass slide) becomes much narrower than the plasmonic resonances and reaches a few meV. Plasmonic NDs have been used in a number of experimental studies due to their directional far-field emission and giant field enhancement achieved in the narrow gap between the nanoparticles [43, 53– 58]. In order to obtain a highly radiative plasmonic mode in the near-infrared (NIR) region, far detuned from the fluorophore ZPL, we further utilize the interaction between the gold ND and a metallic substrate to form a nanodimer-on-mirror (NDoM) structure [59–64], whose optical properties are compared to those of the nanodimer-on-glass (NDoG). These experimental improvements allow us to observe well-resolved and pronounced spectral reshaping of single fluorophore emission. The fluorophore is seen to feed into the far detuned NIR plasmonic mode of NDoM, which does not spectrally overlap with the bare fluorophore emission. This new NIR peak shows intensity fluctuations that follow the blinking dynamics of the fluorophore, while its wavelength and linewidth are stable over time, contrary to the wandering nature of the ZPL and associated vibronic shoulders.

We show that feeding into the NIR mode of the NDoM at low temperature is made possible by a combination of giant Purcell factor and Lamb shift. The Purcell effect leads to an inferred luminescence lifetime of only few tens of femtoseconds that causes lifetime-broadening of the fluorophore emission and permits spectral overlap with the far detuned NIR plasmonic mode. The Lamb shift is responsible for pronounced red-shifts (tens of meV) of the ZPL emission and works in synergy with the Purcell broadening to feed emission into the NIR mode. While our results are quantitatively well reproduced by electromagnetic simulations of the photonic LDOS [65], they also point out to the need for a more accurate theoretical description of the properties of intrinsic dephasing of molecular vibronic transitions, at the microscopic level, to fully capture the new regime of light-nanomatter interaction now accessible with nanocavity molecular emitters, where electronic, photonic and vibrational degrees of freedom all interact strongly with each other [66–69]. Our findings evidence the open questions and challenges in the control and understanding of nanocavity-coupled molecular emitters and may have broad implications for nanoplasmonic approaches to quantum optics and cavity QED [5, 47, 70–72].

## RESULTS AND DISCUSSION

With the help of rectangular DNA-origami structures (60 x 50 nm$^2$) a single ATTO 590 molecule (the fluorophore) is precisely positioned between two Au nanospheres (60 nm nominal diameter) forming the plasmonic nanocavity with a gap width $g \approx 4 - 5$ nm (as expected from the total DNA origami thickness), see Fig. S1. Following their colloidal synthesis in solution (see Methods), NDs are deposited either on glass (NDoG; Fig. 1a) or on a template-stripped gold substrate acting as the mirror (NDoM; Fig. 1 b). The ND is separated from the gold or glass substrate by single stranded DNA sequences that are used to functionalize the Au nanoparticles with a thickness of ≈ 1 to 1.5 nm. We utilize two control samples. The first consists of the same DNA origami containing a single fluorophore attached to a single Au nanoparticle and deposited on glass (NPoG), while the second consists of the fluorophore containing DNA origami deposited directly on a glass substrate.

The plasmonic resonances of individual nanocavities containing single fluorophores are investigated via dark-field scattering spectroscopy (Fig. 1c). By dispersing nanodimers on the substrate with very low surface density, we ensure that the scattering signal arises from a single nanostructure, as confirmed by electron microscopy on a subset of points. The optical response of this configuration is calculated using electromagnetic simulations (Methods and Supplementary Information). The measured (top row) and calculated (bottom row) spectra in Fig. 1c show that the dominant radiative mode of the NDoG lies in the visible region ($\sim 620 - 670$ nm), which overlaps well with the fluorophore ZPL ($\lambda_{em} \simeq 622$ nm at 300 K). The NDoG shows strong scattering only under $s$-polarized illumination (labeled with respect to the substrate, black lines in Fig. 1c), corresponding to the dipolar bonding dimer plasmon (BDP) mode along the dimer axis (see charge and field distribution in Fig. 1d).

In the NDoM configuration the BDP mode is strongly redshifted towards the NIR region (wavelength $\geq 800$ nm) due to the interaction with the metallic substrate, as shown experimentally and by calculations (Fig. 1c, right column). This redshifted mode can be excited by $s$- and $p$-polarized illumination and is characterized by strong surface charge concentration in the three gaps forming the structure (between the two spherical nanoparticles and between the dimer and the gold substrate), as illustrated by the corresponding

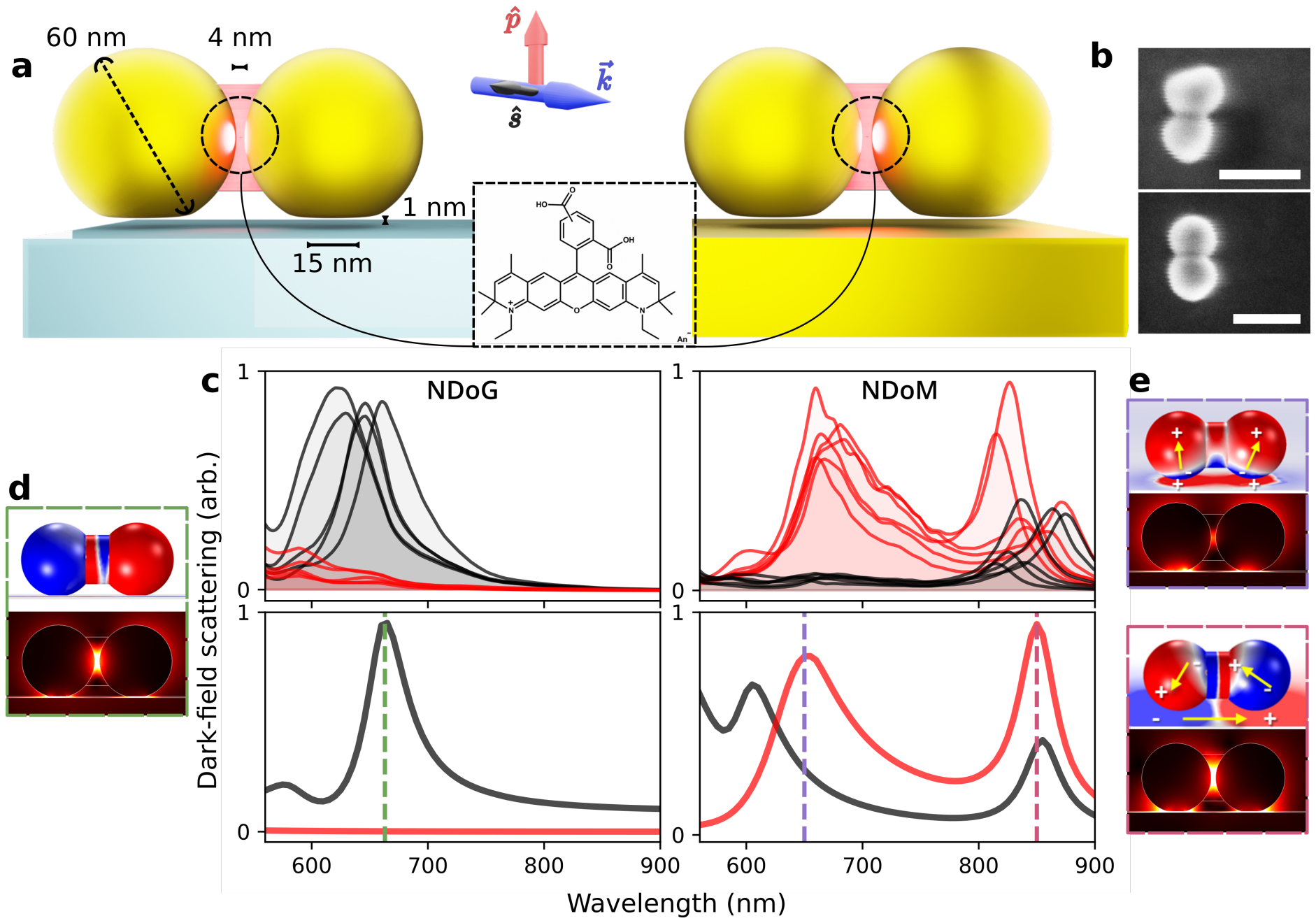

**FIG. 1**. *Optical darkfield characterization of nanocavitites*. **a** Illustration of NDoG (left) and NDoM (right) nanocavities. Relevant dimensions are shown, along with blue arrow indicating the direction of white light used for polarization resolved darkfield scattering spectroscopy. The molecular structure of the ATTO 590 dye positioned in the gap between the two nanoparticles is shown as an inset. **b** Electron microscopy images of single NDoM nanocavities, the scalebar corresponds to 100 nm. **c** Experimentally obtained spectra from individual nano-cavities at room temperature (top row) compared to the results of simulations using the geometry outlined in a (bottom row). Red lines correspond to $p$-polarized excitation, black lines to $s$-polarized. In both experiments and simulations, an excitation angle of $81^{\circ}$ and collection NA of 0.81 is used. **d** Simulated charge (top) and near-field distribution (bottom) for the dominant radiative NDoG mode, at the wavelength indicated by the green dashed lines in **c**. **e** Simulated charge and near-field distributions for NDoM modes, at the wavelengths indicated by the violet and red dashed lines in **c**.

charge and field distributions in Fig. 1e.

The NDoM spectrum also shows a mode in the visible (VIS) domain (630 – 700 nm, Fig. 1c), of a different nature as compared to the main radiative BDP mode of the NDoG cavity. This mode can only be excited by $p$-polarized excitation and is characterized by strong surface charge and field enhancement at the gaps between the nanodimer and the metallic substrate, but not in the gap between the two spherical nanoparticles forming the dimer, where the fluorophore is to be located (the relatively strong local field in these spots in Fig. 1e is due to contributions from other higher order plasmonic modes, which are not revealed in the dark field scattering spectra). The difference in field strength at the fluorophore position compared to the NIR mode is due to different field symmetry at the ND-substrate gaps. In the VIS mode, the orientation of the fields is the same in both gaps, but it is opposite in the NIR mode. With generality, modes exhibiting large field concentration at the gap between nanoparticles (BDP mode in NDoG and NIR mode in NDoM) enable strong radiative enhancement of a dipolar source placed at that position. This enhancement is comparatively weak for the NDoM mode at visible frequencies, but in this case a large enhancement of the decay rate is still feasible as a consequence of the coupling with weakly radiative higher order plasmonic modes typically responsible for quenching (see below).

These general spectral trends can be observed in all the measured particles (Fig. S2), and are correctly reproduced by numerical simulations of the optical response (See Fig. S3 for details on the geometry and Fig. S4a for simulation results).The large variation in NIR mode position in experiments may arise from the variation in nanoparticle shape and DNA coverage, as corroborated by simulations (Fig. S4a). We note that another strategy to redshift the NDoG resonance would be to shrink the inter-particle distance [42]; however, narrower gaps are challenging to achieve using the DNA origami approach, while they could also lead to stronger nonradiative quenching of fluorophore emission [73]. In addition, we will show that our dual-mode strategy allows for maintaining high total LDOS at the ZPL frequency together with the highly radiative NIR mode.

We then perform fluorescence spectroscopy on individual structures at room and cryogenic temperature to study light-matter interaction at the single molecule level (see Methods for the apparatus). All samples containing ATTO 590 fluorophore are measured using an excitation laser at 590 nm, and a long-pass filter that allows us to collect fluorescence beyond 600 nm. The bare ATTO 590 fluorophore (i.e. in the absence of ND) has a broad emission linewidth at room temperature with a zero-phonon line (ZPL) at $\lambda_{em} \simeq 622$ nm (Fig. 2a, blue lines), along with a shoulder corresponding to vibronic sidebands. Upon measuring the fluorescence from bare individual molecules at $T = 4$ K bath temperature, we observe much narrower ($\sim 10$ meV) and asymmetric emission spectra because of the reduced thermal occupancy of vibrational modes and phonon bath [74], as shown with blue lines in Fig. 2b and blue crosses in Fig. 2c (see Fig. S10 for more spectra). Results from NPoGs are shown with golden diamonds in Fig. 2c (see Fig. S11 for example spectra). These structures present a smaller Purcell factor than NDoG and NDoM (2-3 times lower, see Fig. S7), but a similar Lamb shift is expected due to the interaction of the emitter with its image dipole in the nearby metal.

The first major finding of our experimental study is illustrated by the red curves in Fig. 2a,b, corresponding again to measurements at $T = 300$ K (Fig. 2a) and $T = 4$ K (fig. 2b). When a fluorophore is embedded in an NDoM cavity, we observe not only a hundred-fold enhancement of spectrally integrated emission intensity (Fig. S8a) – consistent with an increased excitation rate – and even larger enhancement of total photon budget (Fig. S8b) – consistent with an increased decay rate [75, 76] – but also a dramatic reshaping of the emission spectrum with two main features: (i) a significantly broadened and redshifted ZPL (also seen for fluorophores in NDoG, shown in Fig. S12 ), and (ii) the appearance of a new strong peak around 750 to 800 nm, detuned from the original ZPL by ten to hundred times the main emission peak linewidth at $T = 4$ K. To our knowledge, such a pronounced reshaping of single-molecule emission has not been reported before.

A summary of the main spectral features from the set of investigated individual emitters is presented in Fig. 2c, extracted from time series of single particles measured at $T = 4$ K until photobleaching of the fluorophore. From each time series, the position and linewidth of the ZPL and NIR emissions are extracted and the mean value is represented by a point in Fig. 2c. We keep using the term ZPL to denote, in this context, the highest energy emission peak in its entirety. Error bars are obtained by calculating the standard deviation over each measurement, while the shaded regions are only guides to the eye. A large spread of linewidths is observed for the fluorophore ZPL in NDoG and NDoM nanocavitites. This is assigned to the sensitive dependence of the Purcell factor and the Lamb shift on the exact geometry of the dimer (Figs. S5b), on the position (Fig. S6) as well as the orientation of the fluorophore (Fig. S5a), as will be developed below.

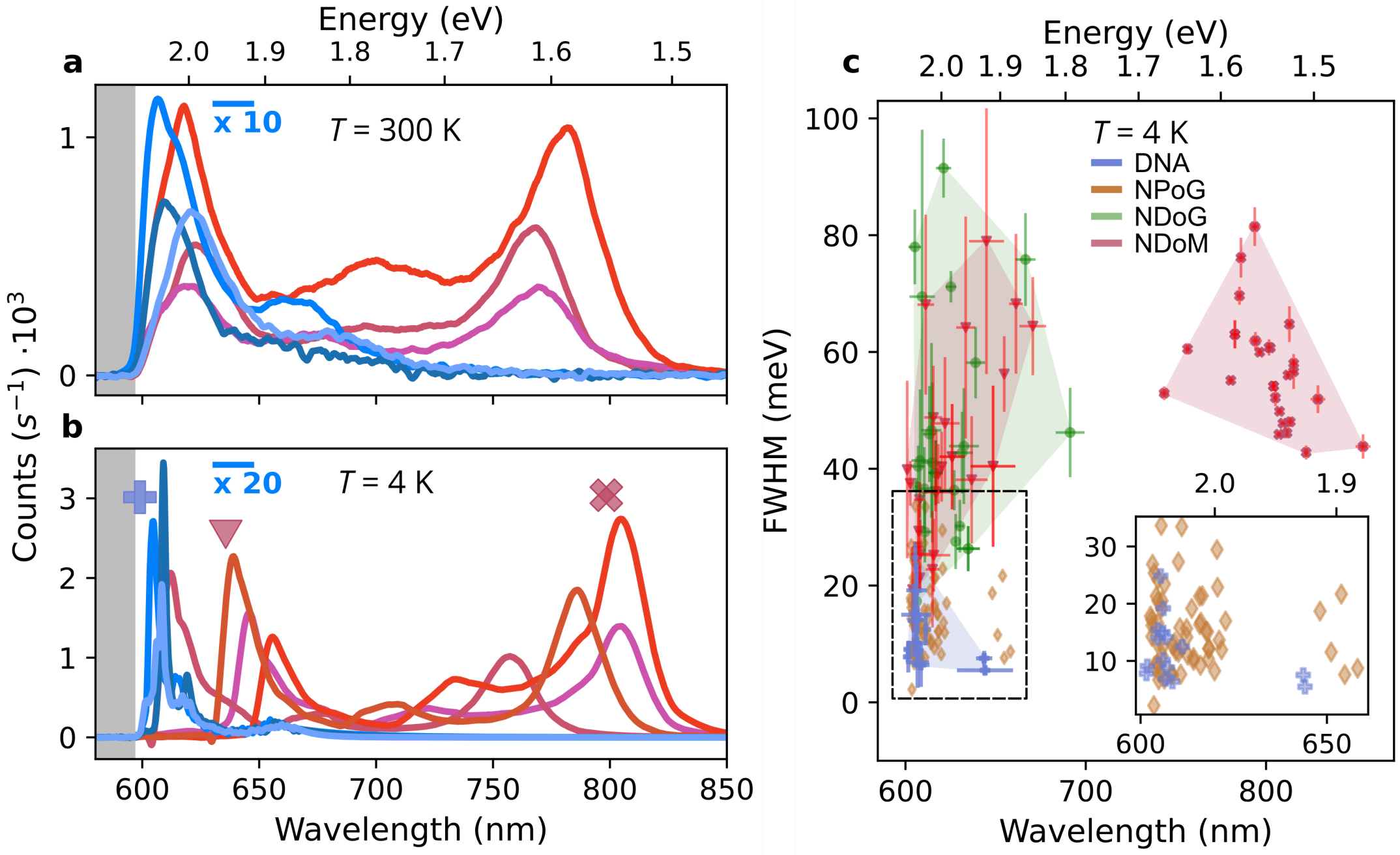

**FIG. 2**. *Spectral reshaping at ambient and cryogenic temperatures.* Typical experimental emission spectra from single molecules in their DNA scaffold deposited on glass (blue curves) or embedded in NDoM cavities (shades of red) at room **a** and cryogenic **b** temperature $T$. A longpass filter cuts the emission below 600 nm. All spectra are measured under 590 nm excitation at 10 $\mu$W incident power (continuous wave) focused through a 0.82 numerical aperture with an acquisition time of 1 s (see Methods for further details). **c** Summary of emission linewidths and positions of emission peaks at $T$ = 4 K. The various groups are plotted with different markers, as indicated in **b** and in the legend. For fluorophores in NDoM (red symbols) the ZPL and far detuned NIR emission are analysed separately. The marginals of the data shown in **c** are presented in S9. Shaded areas are guides to the eye.

Before modelling the reshaped, Purcell-enhanced emission, we study the spectral and intensity fluctuations (conventionally referred to as fluorescence wandering and blinking, respectively) from the three types of samples at low temperature. These fluctuations are illustrated by a few emission time series in Fig. 3a-c. For each measured emitter, the intensity, peak position and full-width at half maximum (FWHM) are extracted frame by frame, while separately treating the ZPL (at visible wavelengths) and cavity-enabled NIR emission for NDoMs. These quantities are then used to compute their respective levels of fluctuation over time, expressed through a boxplot in Fig. S13b, and also study correlations from individual traces as discussed later. In these measurements, we benefit from the prolonged duration (several minutes) before photobleaching of fluorophores, which is a combined effect of cryogenic temperature and plasmon-accelerated decay rate [75, 77].

On the one hand, we observe similar levels of relative intensity fluctuations across all sample types, which contrasts with earlier observations on colloidal quantum dots coupled to metallic structures [78–80] where faster decay has been reported to suppress blinking and stabilize emission intensity [81]. On the other hand, the observed spectral fluctuations of the ZPL are significantly enhanced when embedding the fluorophore in a nanocavity, be it an NDoG or NDoM. The magnitude of wandering of the ZPL is 4 or 5 times more pronounced on average compared to fluorophores in DNA on glass, with some nanocavities featuring spectral excursions larger than 30 meV (S13b). As illustrated by the spectra shown in Fig. 3d and summarized in Fig. 3e, when the ZPL wavelength shifts to longer wavelengths, the NIR emission intensity increases. The full traces are shown in Fig. S16.

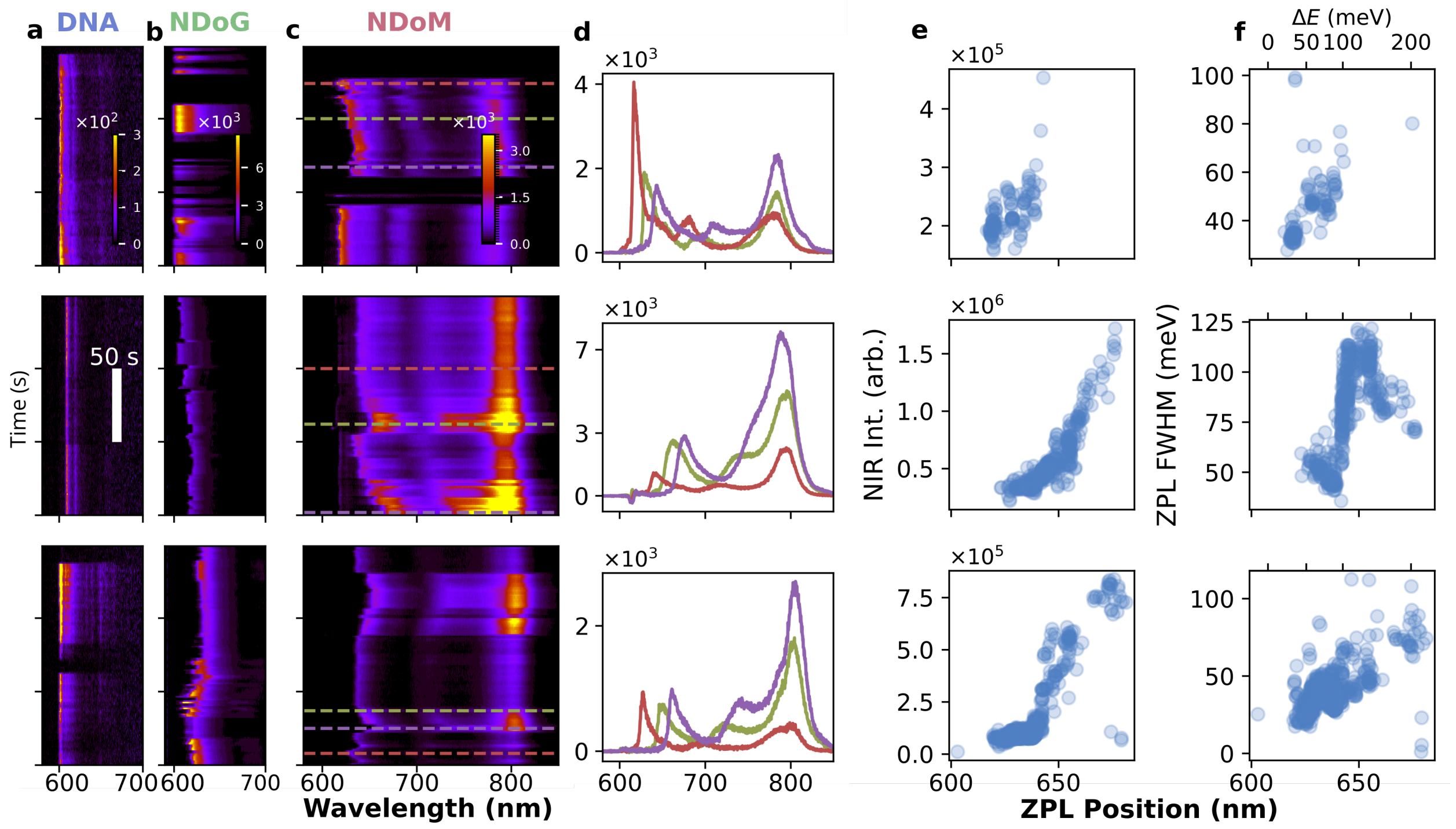

**FIG. 3.** *Spectral and intensity fluctuations.* **a-c** Examples of typical time series of fluorescence spectra acquired from the three sample families presented in Fig. 2**c**. The white bar in **a** indicates the common timescale. The full time-series can be found in Fig. S13. **d** NDoM spectra from the time points marked by lines in the timeseries in **c**, highlighting the correlation between the ZPL position, NIR mode intensity and ZPL linewidth. **e, f** Correlation plots of the ZLP position versus the emitted intensity at the NIR region (integrated from 700 - 900 nm, **e**), and versus the FWHM (**f**), as extracted from each frame of the NDoM timeseries shown in **c**.

In contrast to the wandering fluorophore ZPL, the NIR emission shows very stable center wavelength and linewidth (S13 b), both governed by the plasmonic mode. After the fluorophore has bleached, the much weaker collected signal originates from the intrinsic gold luminescence from the nanocavity, whose spectrum can be used to accurately infer the plasmonic resonant modes [82–85]. This faint post-bleaching gold luminescence shows a peak at the exact same position as the detuned NIR fluorophore emission (Fig. S14), confirming the presence of the corresponding plasmonic mode at the single particle level. All these observations support that the NIR emission is a consequence of the nanocavity mode being fed by the fluorophore emission.

Another interesting correlation can be observed in Fig. 3f and Fig. S15: when the ZPL linewidth broadens, the ZPL emission typically red-shifts. This correlation is consistent with both effects having a common cause, which we attribute to either a transient change in the effective molecule – gold distance, or a change in molecular orientation, possibly resulting from instabilities in the neighboring gold surface configuration [86]. We indeed show in Fig. S6 and Fig. S5a that decreasing the molecule – gold distance, or changing the dipole orientation in the simulation strongly affects both the Purcell broadening and the Lamb-shift. We describe in more details below how this hypothesis is corroborated by theoretical modeling and tested by additional experiments with a different fluorophore (ATTO 643).

We begin by showing with electromagnetic modelling of the system that the Purcell effect and Lamb shift together with the presence of the NIR cavity mode can account for the main experimental features: the pronounced ZPL broadening, the transfer of emission intensity to the NIR peak, and the correlated fluctuations reported in Fig. 3. The NPoG system is a useful control in this regard, since the distance between the fluorophore and gold surface is maintained ($\sim$ 1 nm), while the Purcell factor is significantly reduced. The similarity between the ZPL spectra from ATTO 590 in DNA and in NPoG (inset of Fig. 2c and S11) suggests that no significant change in linewidth is induced by the proximity to the gold surface (Fig. S9), and hence indicates that the broadening and reshaping seen in nano-dimers have a dominantly electromagnetic origin. This is further corroborated by simulations (Fig. S7) that predict a 2-3 times lower Purcell broadening for NPoG compared to NDoM, but a similar Lamb-shift, both aspects being reflected in the ZPL linewidth and wavelength distributions of Fig. 2c.

For NDoG and NDoM cavities, the radiative and total (including non-radiative) decay rates are numerically calculated by modelling the fluorophore as a point-like dipole located in the ND gap, along and parallel to the ND long axis and placed 1 nm away from one nanoparticle (consistent with the DNA origami design); the results are plotted in Fig. 4a (see Fig. S5 for perpendicular configuration). The calculations and other considered geometries are described in more detail in the Methods and Fig. S3. Two features of the NDoM should be highlighted: first, at the ZPL wavelength of ATTO 590 in DNA, which is around 600 nm at 4 K (cf. Fig. 2c), the Purcell Factor (shown in Fig. 4a, green line, and defined as the increase in spontaneous decay rate due to the presence of the ND) reaches $\sim 4 \times 10^4$, while its value is 3 to 4 times smaller around the ZPL of ATTO 643 that will be studied later. Second, the simulation confirms the presence of the NIR mode featuring both large Purcell factor ($\sim 3{\times}10^4$) and high radiative efficiency ($\sim$ 20%) close to 800 nm.

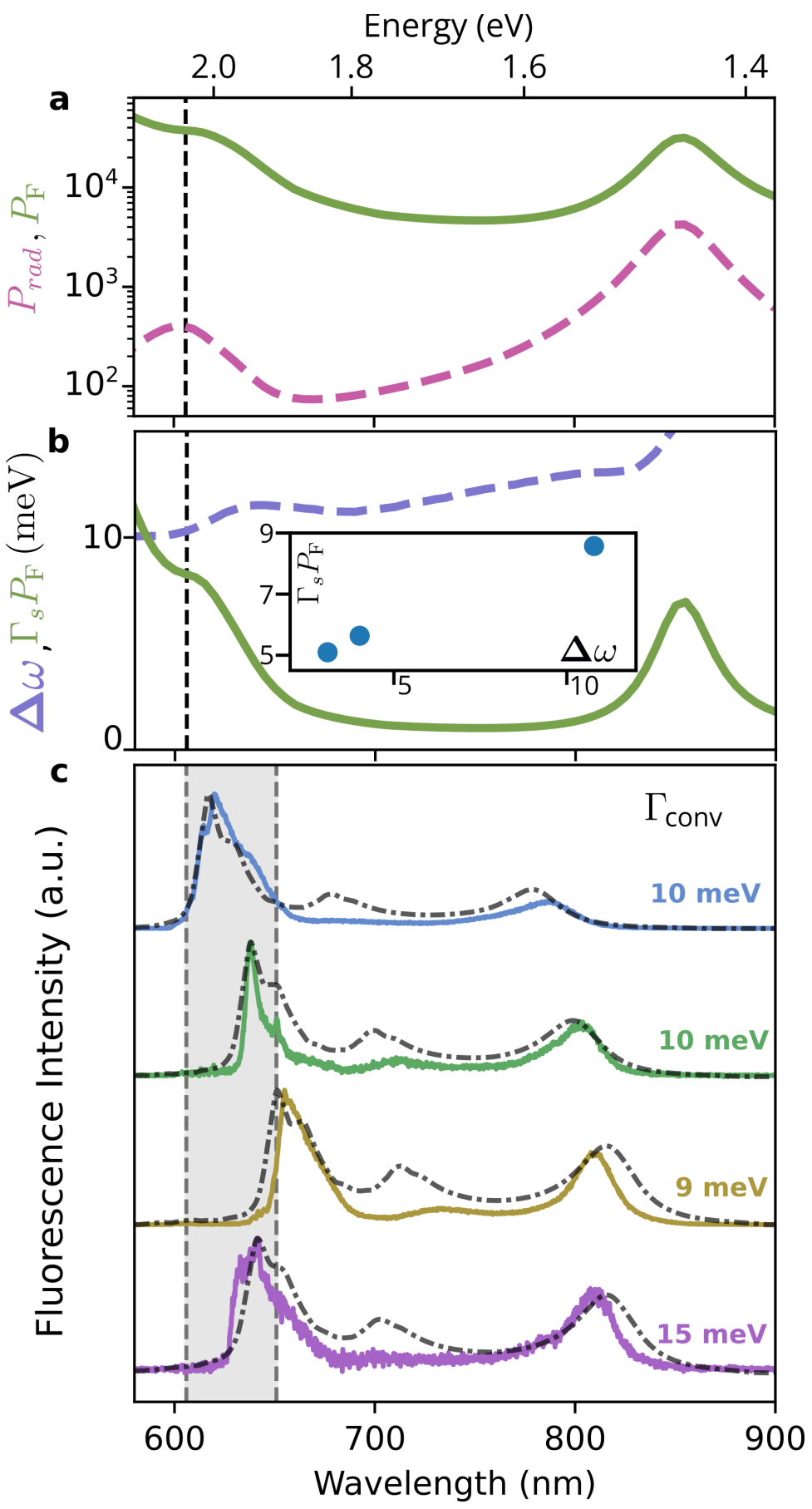

**FIG. 4.** *Modelling luminescence reshaping.* **a**, Simulated radiative ($P_{\text{rad}}$, dashed line) and total ($P_F$, solid line) decay rate enhancement factors as a function of frequency; **b**, Corresponding Purcell broadening ($\Gamma_s \cdot P_F$) and Lamb shift ($\Delta\omega$) in meV, for a spontaneous emission rate ($\Gamma_s$) in the DNA of 0.3 ns$^{-1}$. Electromagnetic simulations are performed using the geometry sketched in Fig. 1**a** with the dipolar emitter placed 1 nm away from one of the Au surfaces. The vertical dashed line in **a, b** indicates the average measured ZPL frequency of ATTO 590 in DNA origami on glass at 4 K. The inset in **b** shows the correlation between the calculated frequency shift and Purcell broadening at the ZPL frequency for various locations of the fluorophore within the nanogap (additional data shown in Fig. S6). **c** Examples of experimental instantaneous fluorescence spectra (solid lines) along with best estimates from the convolution model of eq. (5) (dashed lines). More examples are shown in Fig. S18. While the same $L_{mol}$ is used in all examples, the appropriate $P_{rad}$, correctly matching the NIR mode position (see S4 **b**), is used for each case. The shaded region in **c** shows the range of ZPL energies measured from bare fluorophores in origami.

A naive prediction of the modified fluorophore emission spectrum in NDoM, S($\omega$), is given by the measured bare fluorophore spectrum $L_{\text{mol}}(\omega)$ (examples shown in S10) multiplied by the radiative rate enhancement factor for a dipole in the nanocavity gap ($P_{\text{rad}}(\omega)$, dashed curve in Fig. 4a) according to [23]:

$$S(\omega) \simeq \mathcal{L}_{\text{mol}}(\omega) \cdot P_{\text{rad}}(\omega). \tag{1}$$

This approach, which was used with some success to model room-temperature data [23, 27] (see Fig. S17a), totally fails to reproduce the increased ZPL emission linewidth for a fluorophore embedded in NDoG or NDoM at low temperature (see Fig. S17b); as a result, it also predicts negligible NIR emission for fluorophores in NDoM. We therefore refine the model by accounting for the reduced luminescence lifetime of a fluorophore coupled to a nanocavity, which results in a broadened ZPL. When the emitter is modeled by a two-level system, it's emission spectrum in a reference environment can be modeled by a Lorentzian function. Using the formalism in [87, 88] with the frequency dependent Purcell factor ($P_F(\omega)$) and Lamb shift ($\Delta\omega(\omega)$) inside the cavity, the spectrum is modified as:

$$\mathcal{L}_{\mathrm{mol}}^{\mathrm{cav}}(\omega) = \frac{1}{\left(\omega-\omega_0+\left(\frac{2\omega_0}{\omega+\omega_0}\right)\Delta\omega(\omega)\right)^2+\left(\frac{2\omega_0}{\omega+\omega_0}\right)^2\left(\frac{(\Gamma_{\mathrm{deph}}+\Gamma_{\mathrm{cav}}(\omega))}{2}\right)^2} \quad (2)$$

where $\omega_0$ is the central emission frequency in the absence of a cavity, $\Gamma_{\mathrm{deph}}$ is the pure dephasing rate (that vastly dominates the linewidth of the bare fluorophore) and $\Gamma_{\mathrm{cav}}(\omega) = P_F(\omega)\Gamma_s$ is the Purcell enhanced decay rate (and corresponding broadening) inside the nanocavity, with the spontaneous radiative decay rate of the bare fluorophore $\Gamma_s$ (Fig. S19).

Finally, the photoluminescence signal of the coupled system is calculated by:

$$\mathcal{S}^{\mathrm{cav}}(\omega) = \mathcal{L}_{\mathrm{mol}}^{\mathrm{cav}}(\omega)\cdot P_{\mathrm{rad}}(\omega) \quad (3)$$

In this model all vibrational degrees of freedom of the molecule are phenomenologically included into the pure dephasing rate $\Gamma_{\mathrm{deph}}$. However, as the low temperature fluorescence spectra feature asymmetric ZPL and vibronic sidebands, the two-level approximation must be improved upon. We introduce a heuristic model based on the convolution of the reference fluorophore spectrum with a Lorentzian function that accounts for Purcell broadening through the broadening parameter $\Gamma_{\mathrm{conv}}$, which can be identified with the predicted value of $P_F\Gamma_s$ close to the Lamb-shifted ZPL energy:

$$(4) \quad \mathcal{L}_{\mathrm{mol}}^{\mathrm{conv}}(\omega) = \int \mathcal{L}_{\mathrm{mol}}^{\mathrm{ref}}(\omega-\omega')\,\frac{\Gamma_{\mathrm{conv}}/2}{\pi((\Gamma_{\mathrm{conv}}/2)^2+(\omega')^2)}d\omega'$$

In the following, we choose the spectra of fluorophores coupled to single gold nanoparticles on glass (NPoG) as the reference $L^{\mathrm{ref}}_{\mathrm{mol}}$ because they already contain the effect of the Lamb shift (which is mostly due to the proximity of a metal surface) and possible extraneous effects of the nearby metal on the linewidth, while experiencing negligible lifetime broadening. The expected reshaped spectra is calculated via:

$$\mathcal{S}^{\mathrm{conv}}(\omega) = \mathcal{L}_{\mathrm{mol}}^{\mathrm{conv}}(\omega)\cdot P_{\mathrm{rad}}(\omega) \quad (5)$$

The predictions of this improved heuristic model, shown in Fig. 4, allow us to recover the correct lineshape for the ZPL, while also being able to predict the correct NIR intensity, with values of $\Gamma_{\mathrm{conv}}$ that are comparable to the predicted lifetime broadening through the Purcell effect using a fluorophore spontaneous decay rate ($\Gamma_s$) of 0.3 $\mathrm{ns}^{-1}$, very close to that measured in Fig. S19. Even with our simple assumptions, the Purcell model captures some of the essential features of the system while reproducing experiments with very satisfying accuracy.

To further test our hypothesis regarding the importance of Purcell broadening, we investigate a second set of nanocavities that host a different fluorophore (ATTO 643, Fig. 5 ). As seen from the bare fluorophore spectra (Fig.

S10), ATTO 643 ZPL (avergage ZPL emission wavelength $\sim 646$ nm at 4K) is red-shifted compared to ATTO-590 ($\sim$ 606 nm). However, even though ATTO 643 is tuned closer to the NIR mode, the observed NIR intensity is weaker on average (Fig. 5c), which we attribute to lower Purcell broadening. Indeed, by choosing a subset of all measured samples that may have a similar geometry (based on their plasmonic NIR resonances, gray strip in Fig. 5b), we find that the ZPL linewidth is significantly narrower for ATTO 643, as shown in Fig. 5a (see also Fig. S20a) – which is in line with the NDoM simulations predicting a lower total LDOS around 650 nm compared to 600 nm (Fig. 4a). Note that this trend is robust and visible without any post-selection (Fig. S21). Altogether, the fluorophore that is detuned further from the NIR mode is feeding it more efficiently, in accordance with our model that accounts for giant Purcell broadening and Lamb shift at the single molecule level.

While this model seems to capture the essential physics at play, several other mechanisms will require future investigations. In particular, the predicted total Purcell factor corresponds to a shortening of emission lifetime below 100 fs. Under such rapid decay, Kasha's rule, which states that emission always occurs from the lowest vibrational level of the excited electronic state, could be violated [27, 89]. Since the Frank-Condon factors from vibrationally excited levels may substantially differ from those involving the lowest level, it is unclear what the actual spectrum would be in this case (we would naively expect emission at higher energies, which does not seem to occur). Testing this hypothesis may require ultrafast transient measurements with temporal resolution of few tens of femtosecond on single nanocavities [90, 91]. On the theoretical side, there is a need for solving a model that non-perturbatively accounts for the coupling of the electronic states to the molecular vibrations and to the dynamically screened local electromagnetic field, while properly describing all dissipation channels. A few recent works can serve as a basis for this extension [66, 92]. It is possible that the new terms coming from the Purcell effect acting on vibronic states can more accurately reproduce the experimental spectra.

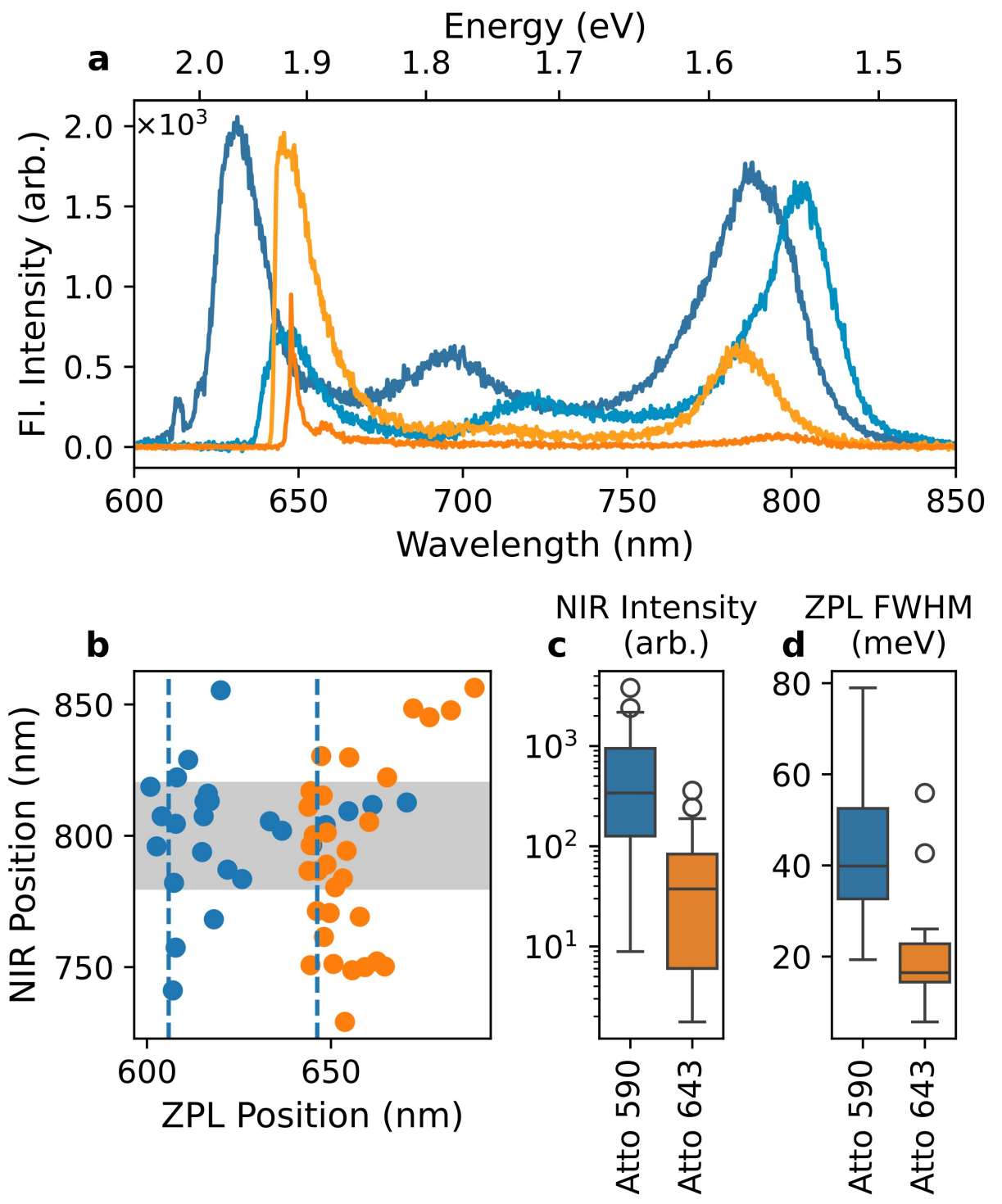

**FIG. 5**. *Experimental comparison of Purcell broadening and NIR emission for two different fluorophores.* **a** Examples of typical fluorescence spectra from NDoMs containing individual ATTO 590 (shades of blue) and ATTO 643 (shades of orange) fluorophores. **b** Scatter plots showing the measured ZPL and NIR emission wavelengths for the two sets of NDoM samples containing different fluorophores. **c** Statistics of NIR emission and d ZPL FWHM of the selected subset of nanocavities that have an NIR mode wavelength between 780 − 820nm, indicated by the shaded region in b.

Furthermore, as presented in Fig. S22, a few rare instances of NDoMs show extreme reshaping events in the emission spectra. Occasionally, the ZPL emission becomes very weak or even disappears while the emission through the NIR mode is further strengthened. This indicates that there may be other microscopic mechanisms at the single-molecule level which are not captured in our model and may lead to quasi-complete transfer of energy from the ZPL to the NIR cavity mode. Such mechanisms may include the formation of picocavities (as observed under much higher laser powers in [86]), the possible activation of spin-forbidden transitions by magnetic modes [93–96], or may be related to increased charge noise in the immediate vicinity of the fluorophore [97–103] due to the presence of the metallic nanoparticles.

In summary, we experimentally studied single molecule fluorescence reshaping due to the coupling with a plasmonic nanocavity at low temperature. By investigating bare fluorophores, as well as fluorophores embedded in two types of plasmonic nanocavities (NDoG and NDoM), and consistently with electromagnetic model calculations, we could interpret the pronounced emission reshaping as a consequence of giant Purcell enhancement and associated Lamb shift, which together enable emission feeding into far detuned cavity modes at NIR wavelengths (up to 850 nm). Our results may have important consequences for the design of broadband single photon sources and applications in quantum communication based on single quantum emitters. For example, they suggest an alternative method for creating NIR emitters, based on organic fluorophores, with high quantum yield and photostability. The coupling of these emitters with plasmonic structures can lead to the development of brighter sources with tailored spectra. Importantly, our analysis suggests that we observe lifetime-limited emission width from fluorophores whose linewidth is otherwise vastly dominated by pure dephasing under usual conditions, which opens new prospects for the creation of indistinguishable single photons from commercial dye molecules. [104–106]

## ACKNOWLEDGEMENT

This work received funding from the European Union's Horizon 2020 research and innovation program under Grant Agreement No. 820196 (ERC CoG QTONE). C.G. also acknowledges the support from the Swiss National Science Foundation (project numbers 170684 and 198898). M. Z.-H., R. E. and J.A. acknowledge funding from grant PID2022-139579NB-I00 funded by
MCIN/AEI/10.13039/501100011033 and by "ERDF A way of making Europe", from grant no. IT 152622 from the Basque Government for consolidated groups of the Basque University, as well as by the European Union (NextGenerationEU) through the Complementary Plans (Grant No. PRTR-C17.I1) promoted by the Ministry of Science and Innovation within the Recovery, Transformation, and Resilience Plan of Spain, and part of the activities of the IKUR strategy of the Department of Education promoted by the Department of Education of the Basque Government. G.P.A. acknowledges support from the Swiss National Science Foundation (200021_184687) and from the National Center of Competence in Research Bio-Inspired Materials NCCR (51NF40_182881).

---

* These two authors contributed equally

† ruben.esteban@ehu.eus

‡ guillermo.acuna@unifr.ch §
chris.galland@epfl.ch

# Giant Purcell broadening and Lamb shift for DNA-assembled near-infrared quantum emitters

## METHODS

### Sample Fabrication

Gold NDs were assembled by means of a rectangular DNA origami [43]. The DNA origami structure was designed using CaDNAno software [107] and visualized for twist correction using CanDo [108]. It consists of a square lattice design that folds into a 2-layers sheet (2LS) of size $\sim$ 60 nm × 50 nm × 5 nm with one ATTO 594 or ATTO 643 fluorophore in the center. 8 handles of (Poly-A15) have been extended from each side of the structure so that it could bind two 60 nm gold nanoparticles surface-functionalized with Poly-T8/18 (Biomers GmbH). The handles have been designed so that the attached nanoparticles are located in the center of the structure with a single fluorophore between them (see Fig. S1 for a schematic and a list of staples). For the structures assembling a single nanoparticle, a DNA origami structure containing handles only on one side (corresponding to the fluorophore side) was assembled.

DNA origami folding was carried out by mixing M13mp18 scaffold (20 nM) with unmodified (200 nM) and modified staples (handles and fluorophore, 2000 nM) in a 1xTAE buffer containing 12 mM $MgCl_2$. Unmodified staples were purchased from Integrated DNA Technologies IDT, modified ones from Biomers GmbH. The use of a scaffold-to-staple ratio of 1:10 and 1:100 guarantees complete integration of the staples in the structure, especially in the case of the handle strands and fluorophore. For folding, the solution was firstly heated up to 75$^\circ$C and then ramped down to 25$^\circ$C at a rate of 1$^\circ$C every 20 min. The folded DNA origami structures were purified from excess staple strands by gel electrophoresis using a 1% agarose gel (LE Agarose, Biozym Scientific GmbH) in a 1xTAE buffer with 12 mM $MgCl_2$ for 2.5 hours at 70 V. The appropriate band containing the targeted 2LS origami was cut out and squeezed using cover slips wrapped in parafilm. The concentration was determined via UV-Vis absorption spectroscopy (Nanodrop).

For binding to DNA origami, gold nanoparticles were functionalized with the mix of Poly-T8 and PolyT18 DNA strands (Biomers GmbH) complementary to Poly-A15 on the surface of DNA origami. Firstly, 125 $\mu$L of thiolated DNA (100 $\mu$L of Poly-T8 and 25 $\mu$L Poly-T18) was activated with 50 $\mu$M TCEP for 1h in order to brake disulfide bonds between SH-DNA strands. Then, activated DNA strands were mixed with 312.5 $\mu$L of 0.1% SDS solution of 60 nm gold nanoparticles of OD 20, adjusted to 1 mM NaCl, and frozen overnight [109]. Afterwards, DNA-functionalized nanoparticles were purified from an excess of DNA strands using gel electrophoresis (1% agarose gel, 2h, 100V). This step also ensures the removal of any self-aggregated dimer formed during the NP functionalization. The concentration of gold nanoparticles was determined via UV-Vis absorption spectroscopy (Nanodrop). The purified 2LS origami was mixed with the purified gold nanoparticles using an excess of five gold nanoparticles per binding site and adding NaCl to a final concentration of 600 mM. After overnight incubation at room temperature, the excess of gold nanoparticles was removed by gel electrophoresis (running for 4.5 hours), and the band containing correctly formed dimers (or the monomer band for the single nanoparticle structures) was extracted as described before.

The fabricated structures are then incubated for 30 minutes for electrostatic immobilization on a dielectric (commercially available glass slides) or gold substrate, rinsed two times with water, and flushed with nitrogen before measuring. Ultra flat gold films are prepared using the template-stripping method as in [110] to ensure reproducible nano-gaps.

## Cryogenic fluorescence spectroscopy

Single particle optical microscopy is performed on a home-built confocal microscope setup. The sample itself is housed within a closed cycle optical cryostat (AttoCube AttoDry 800) and mounted on a 3-axis nano-positioner (AttoCube ANP series) for precisely locating single nanoparticles for measurement. The sample temperature within the cryostat can be precisely controlled between 4 and 300 K using a combination of closed-cycle Helium pumping and resistive heating. A high numerical aperture objective (AttoCube LTAPO 100x 0.82 NA) inside the cryostat is used to tightly focus light from a tunable continuous wave optical parameter oscillator (Hübner C-Wave OPO) onto the sample. Light reflected from the focal spot is re-collimated through the same objective. After blocking the laser line, the remnant signal of interest is focused into a spectrometer (Andor Kymera 193i) with CCD camera (Andor iDus).

Measuring organic fluorophores requires us to tune the laser wavelength to the absorption maximum of the fluorophore and operate at safe excitation power levels. All spectra shown in our report are measured with the excitation wavelength at $590$ nm for ATTO 590 and $633$ nm for ATTO 643, with an excitation power of $10\ \mu$W. Single emitters are located by scanning the sample stage while acquiring signal. When a single emitter is located, time series of spectra are acquired at a rate of $1$ Hz. The length of the time-series is made long enough to capture a bleaching step as well as subsequent background signal from the same spot. Subtracting the average background allows us to extract the spectra emanating from the fluorophore in the focal spot. Only spectra from time-series showing a single bleaching step have been included in this study.

## Room temperature Dark-Field scattering

Dark-field spectroscopy is performed by weakly focusing (Olympus Plan N 4x 0.1NA) a white light source on the sample at a high angle (81$^\circ$) to the normal. The scattered light is then collected through a high-NA objective (Olympus MPLFLN 100x 0.9NA) perpendicular to the sample plane. We use a pulsed supercontinuum white light source (NKT Photonics SuperK Compact) after scrambling its spatial coherence through a spinning diffuser to avoid strong fringes in the scattering spectra. A small percentage of the scattered light is sent to a camera to image the sample while the remainder is sent to a spectrograph (Andor Kymera 193i) with CCD (Andor iDus) to measure the scattering spectra.

## Simulation

All simulations carried out in this work are implemented in the radio-frequency (RF) module of COMSOL Multiphysics [65], which uses the finite element method (FEM) to solve Maxwell's equations in the frequency domain. To mimic the experimental configurations, the simulations in the main text consider two gold spherical nanoparticles (modeled using the permittivity of Au from Johnson and Christy [111]) of 60 nm diameter, separated by a small gap of 4nm. This dimer is deposited on top of a semi-infinite glass of refractive index 1.5 (nanodimer-on-glass NDoG configuration) or a 1.0 nm thick dielectric layer (refractive index 1) placed over a gold substrate (nanodimer-on-mirror NDoM configuration). To represent the DNA origami structure containing the fluorophore, a cylinder of diameter 40 nm and refractive index 2.15 [112] is placed between the two spherical nanoparticles. Other configurations are discussed in the Supplementary Information.

The dark field scattering of the NDoG and NDoM structures is calculated in COMSOL by configuring a background electric field corresponding to a linearly polarized plane wave ($s$- and $p$-polarizations) at an oblique incidence angle of 81$^\circ$ from the surface normal. No fluorophore is considered in these simulations of the scattering. To replicate the experimental collection of light through a high-NA objective, the scattered light is computed by integrating the Poynting vector over a surface area corresponding to a solid angle equivalent to a numerical aperture (NA) of 0.9. The total ($\Gamma_{\rm cav}$) and radiative ($\Gamma_{\rm rad}$) decay rates, as well as the Lamb shift, are computed using as a source an electric point-dipole placed in the nanodimer gap. In the simulations of the main text, the dipole is positioned on, and oriented parallel to the nanodimer's long axis and placed 1 nm away from one of the surface of one nanoparticle. $\Gamma_{\rm cav}$ and the Lamb shift are calculated from the self-interaction Green's function, which gives the electric field (component aligned with the electric dipole) induced by the dipole at the same position of the dipole. $\Gamma_{\rm rad}$ is calculated as the integral of the Poynting vector evaluated in a closed spherical surface surrounding the system at a distance of 1.6 $\mu$m. $\Gamma_{\rm cav}$ and $\Gamma_{\rm rad}$ are then normalized to the same quantities obtained with the dipole immersed in an infinite space made of DNA (refractive index 2.15). The full simulation environment for all calculations consisted of a sphere with a radius of 2000 nm surrounded by a 400-nm-thick perfectly matched layer (PML). Tetrahedral elements were used to mesh all domains, with the maximum mesh element size kept below $\lambda/10$, where $\lambda$ is the wavelength of the excitation. For elements corresponding to the dimers and the DNA cylinder, a finer mesh size was used. Additionally, the dipole source was surrounded by a sphere with the same refractive index of DNA and a diameter equal to the gap size (4 nm) to ensure a sufficiently fine grid in close proximity to the dipole source. The mesh of the sphere surrounding the dipole source was smaller than 1.2 nm. Mesh sizes in the entire set of calculations were systematically verified to ensure convergence was achieved.

SUPPLEMENTARY FIGURES

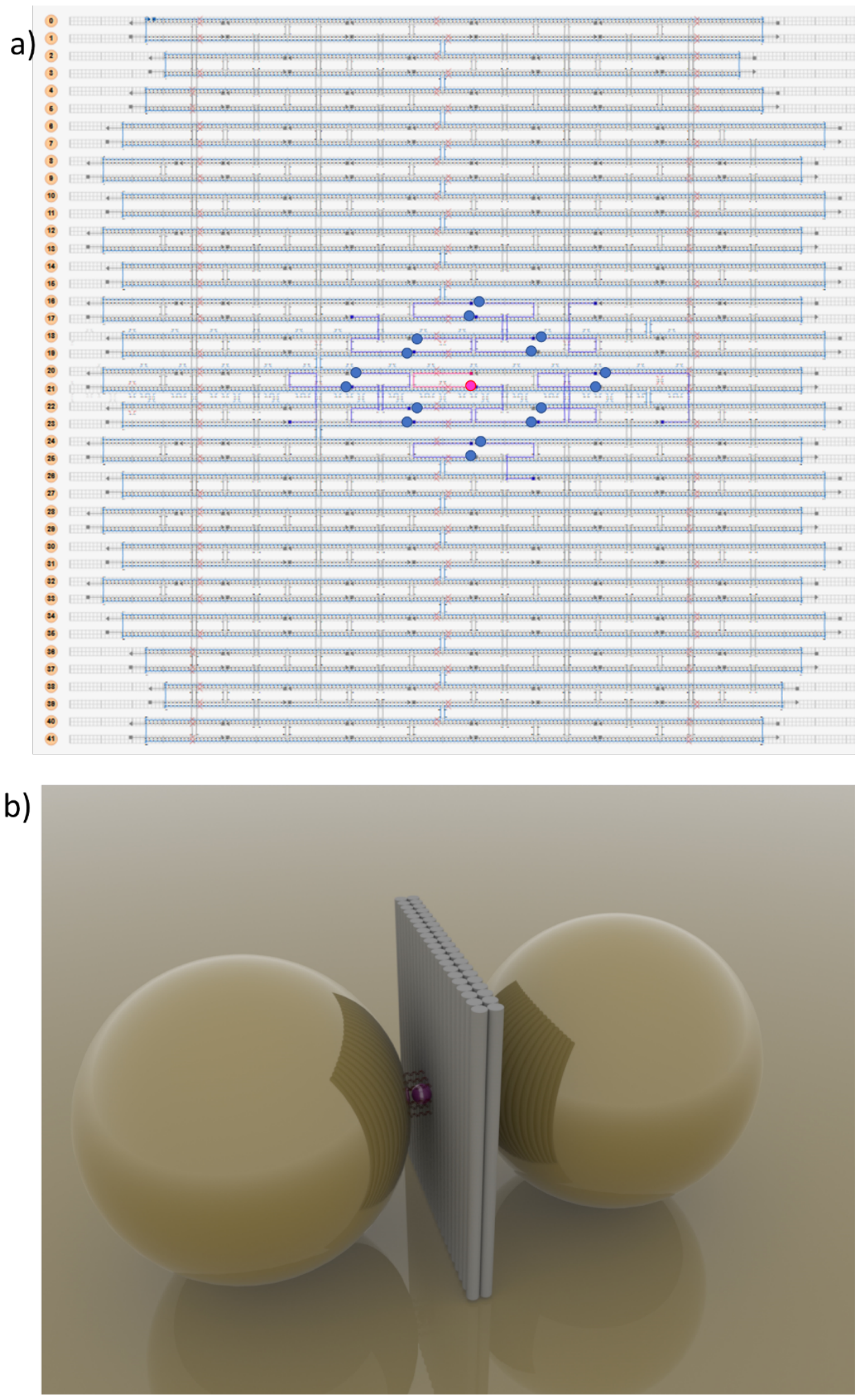

**FIG. S1**. DNA origami sketch. **a** CaDNAno design of the 2LS used to assemble the NDs. The position of the modified staples used for binding Au nanoparticles is represented by blue circles. The position of the fluorophore is indicated by a pink circle. **b** Schematic representation of the obtained gold NDs deposited on a gold substrate. The DNA forms a rectangular structure of size ∼ 60 nm × 50 nm × 5 nm with a single fluorophore off-center in the nanogap. The fluorophore is represented here also by a pink ball between the nanoparticles, and is ∼ 1 nm away from the Au nanoparticle surface.

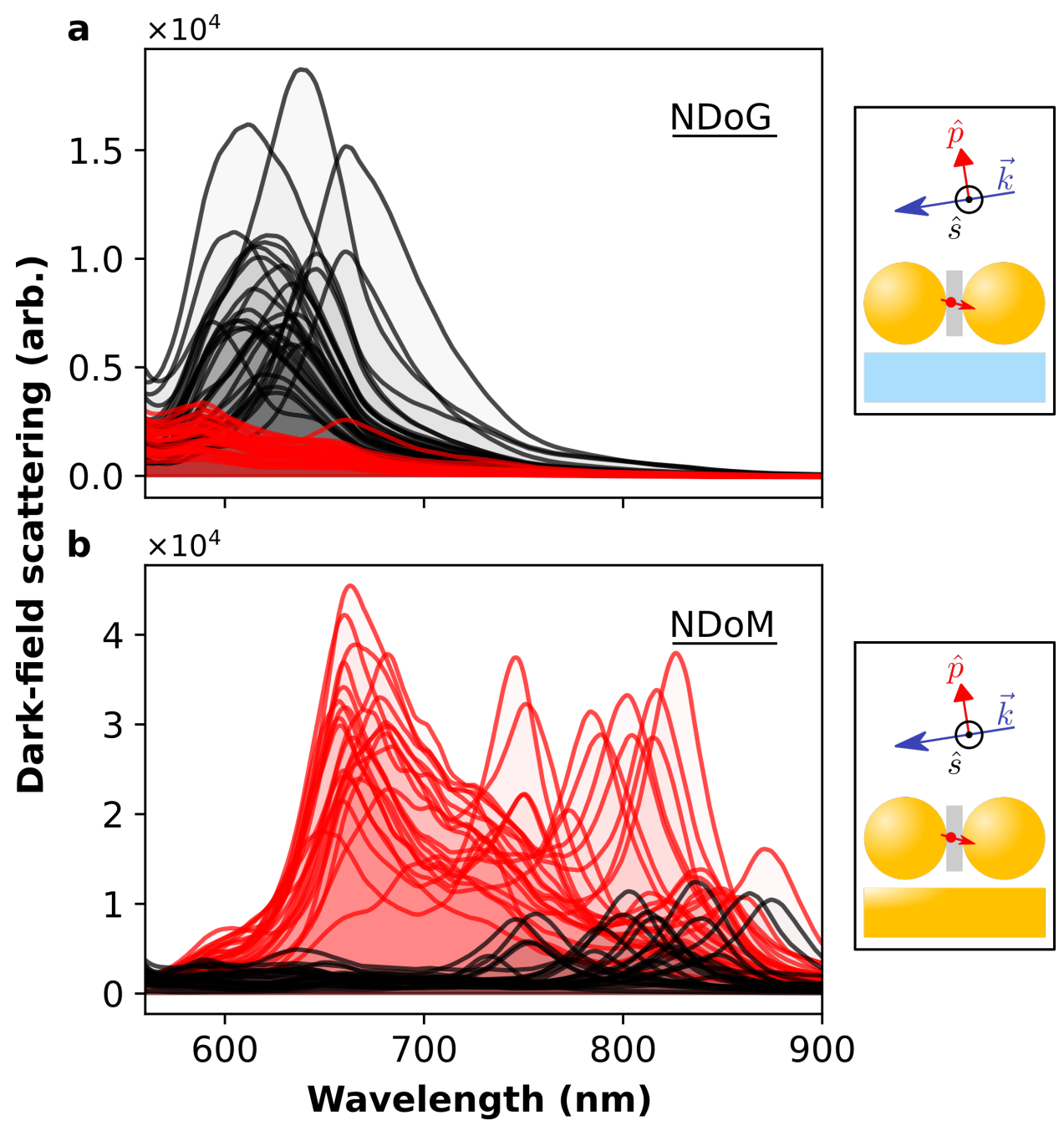

**FIG. S2.** Room-temperature dark-field scattering: Dark-field scattering spectra of individual NDoG **a** and NDoM **b** cavities. Each nanoparticle is excited by $s$- and $p$-polarized white light. An excitation angle of 81° and collection NA of 0.9 are used.

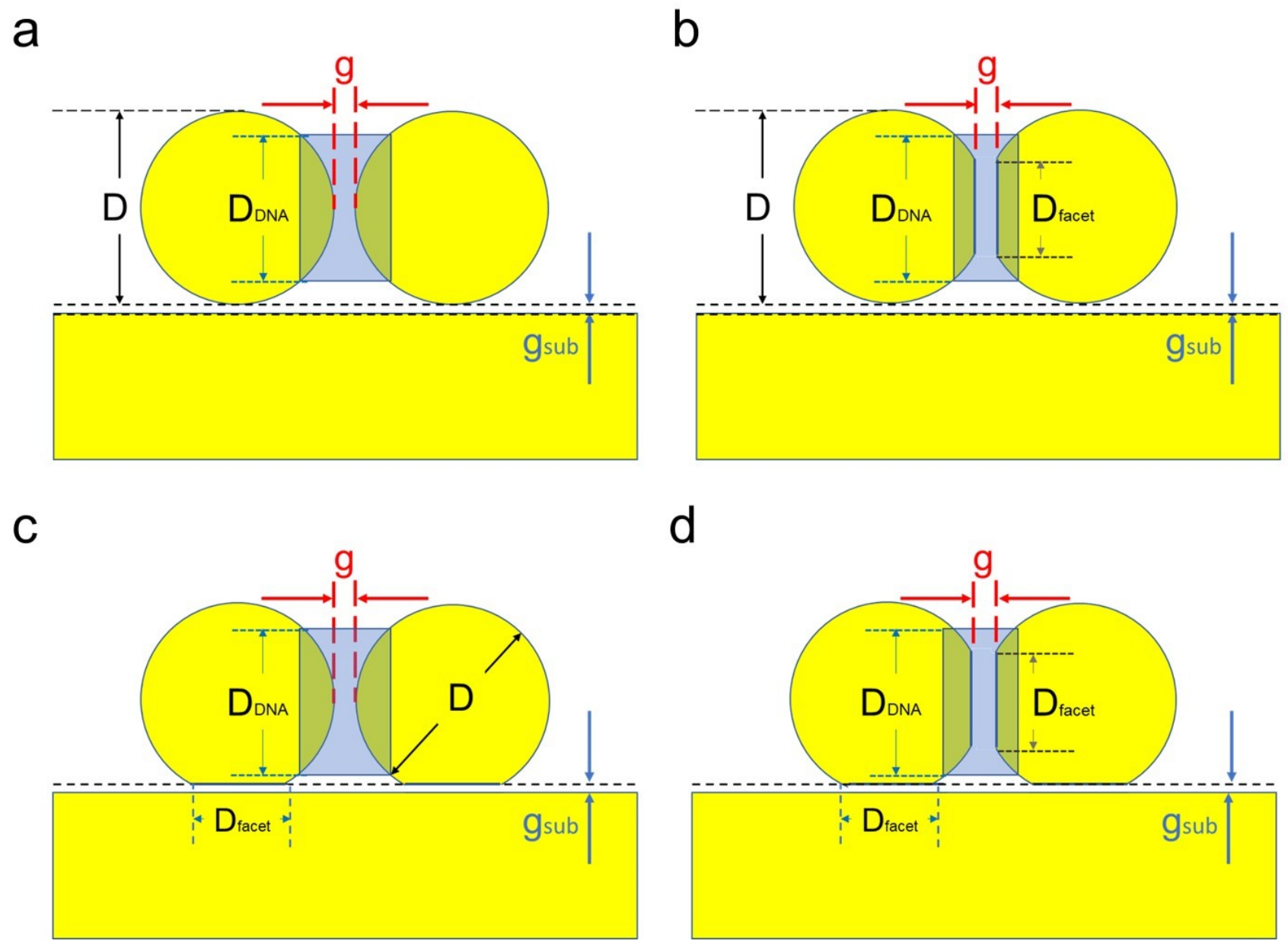

**FIG. S3.** Geometries of the different NanoDimer-on-Mirror (NDoM) configurations used in the main text and in the SI. In all cases, **a** Au nanodimer is situated over **a** Au substrate (the "mirror") forming an interparticle gap between the two nanoparticles and two nanogaps between each of the particles and the substrate. **a** Au spherical nanoparticles with no facets ("No facet" geometry). **b** Au spherical nanoparticles with flat facets in the gap between the nanoparticles ("$f_{top}$" geometry). **c** Au spherical nanoparticles with bottom flat facets in the gap between the nanoparticles and the Au substrate ("$f_{bottom}$" geometry). **d** Au spherical nanoparticles with flat facets in the gap between the nanoparticles and the Au substrate as well as in the interparticle gap ("$f_{top}$ ; $f_{bottom}$" geometry). In all calculations in the main text and in the Supplementary material, the diameter of the gold nanoparticles ($D$), the diameter of the flat facets ($D_{facet}$) and the minimum gap distance ($g$) between the two particles are kept constant at 60 nm, 15 nm, and 4 nm, respectively. The minimum dimer-substrate gap distance ($g_{sub}$) is either 1 nm or 1.5 nm ($g_{sub}$=1 nm in the main text; otherwise indicated in the caption of the Supplementary figures). The DNA origami structure containing the fluorophore is modelled as a cylinder placed between the two spherical nanoparticles (represented by the semi-transparent region) of diameter $D_{DNA}$ = 40 nm and refractive index $n_{DNA}$ = 2.15. The permittivity of the Au spherical nanoparticles and the substrate is adopted from Johnson and Christy [111]. The surrounding medium is vacuum. The geometry of the NanoDimer-on-Glass (NDoG) structure in Fig. 1 of the main text is as in panel c but the dimer is deposited on top of a semi-infinite glass of refractive index 1.5.

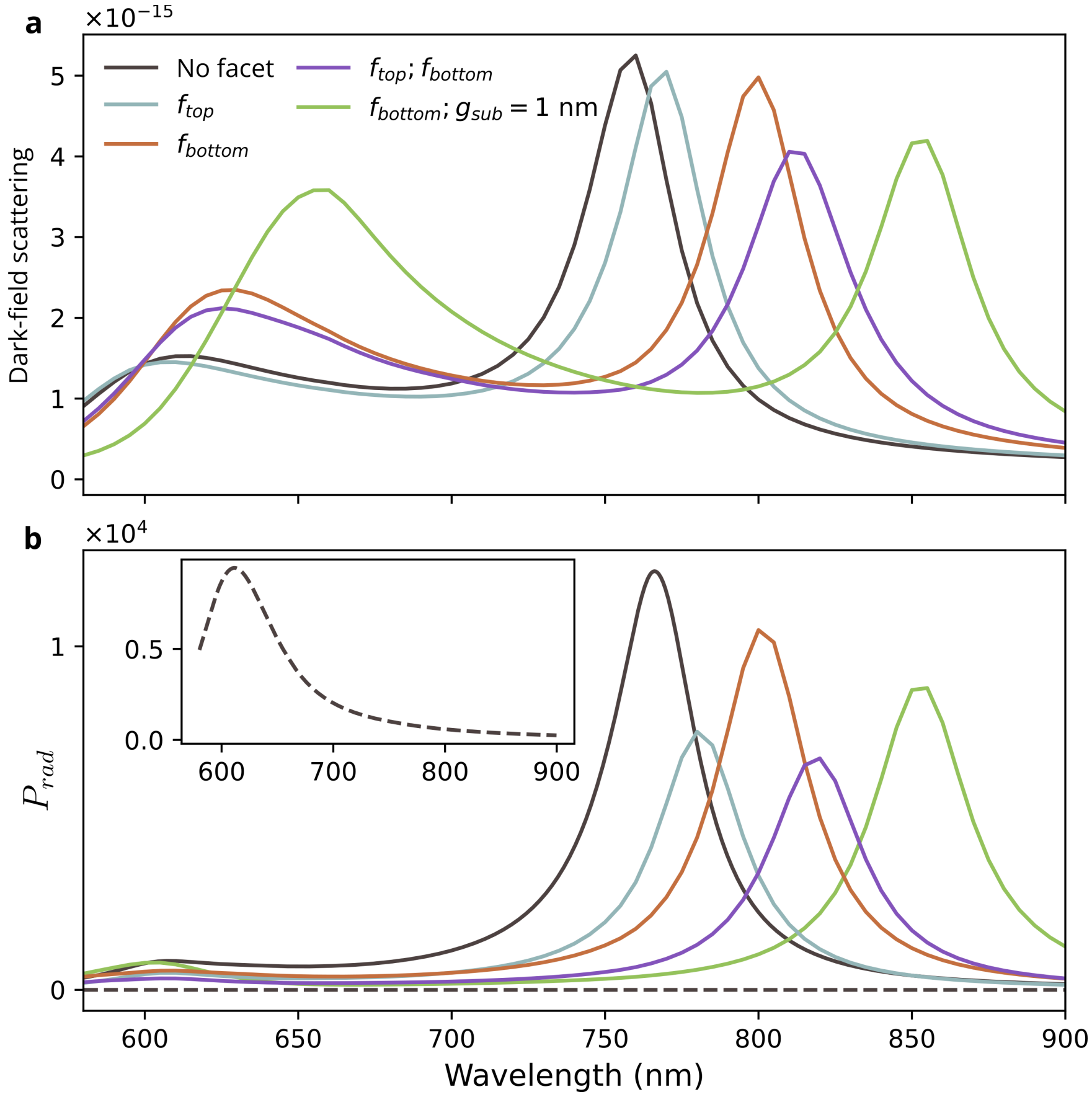

**FIG. S4**. Plasmonic resonances in NDoM cavities. **a** $p$–polarized dark-field scattering spectra calculated for the NDoM geometries in Fig. S3. Black: Au spherical nanoparticles with no facets (Fig. S3a; "No facet") and with $g_{sub} = 1.5$ nm. Gray: Au spherical nanoparticles with flat facets in the gap between the two nanoparticles (Fig. S3b; "$f_{top}$") and with $g_{sub} = 1.5$ nm. Brown, Green: Au spherical nanoparticles with flat facets in the gap between the nanoparticles and the Au substrate (Fig. S3c; "$f_{bottom}$") and with (green) $g_{sub} = 1nm$ or (brown) $g_{sub} = 1.5$ nm . Violet: Au spherical nanoparticles with flat facets in the gap between the nanoparticles and the Au substrate as well as in the interparticle gap (Fig. S3d; "$f_{top}$

; $f_{bottom}$") and with $g_{sub} = 1.5$ nm. The NDoM is illuminated by a plane wave at an excitation angle of $81^{\circ}$, and the collection numerical aperture is $NA = 0.9$. These COMSOL simulations rationalize the variation in NIR mode position seen in experiments, which is very sensitive to the geometry. **b** Simulated radiative rate enhancement for the same nanocavity geometries with a dipole placed at the center of the gap between the two nanoparticles and oriented along the dimer axis (parallel to substrate surface). The radiative efficiency is highest at the wavelengths of the NIR plasmonic mode. The inset shows results for the spherical geometry, with the dipole oriented perpendicular to the dimer axis, where no radiative enhancement is predicted.

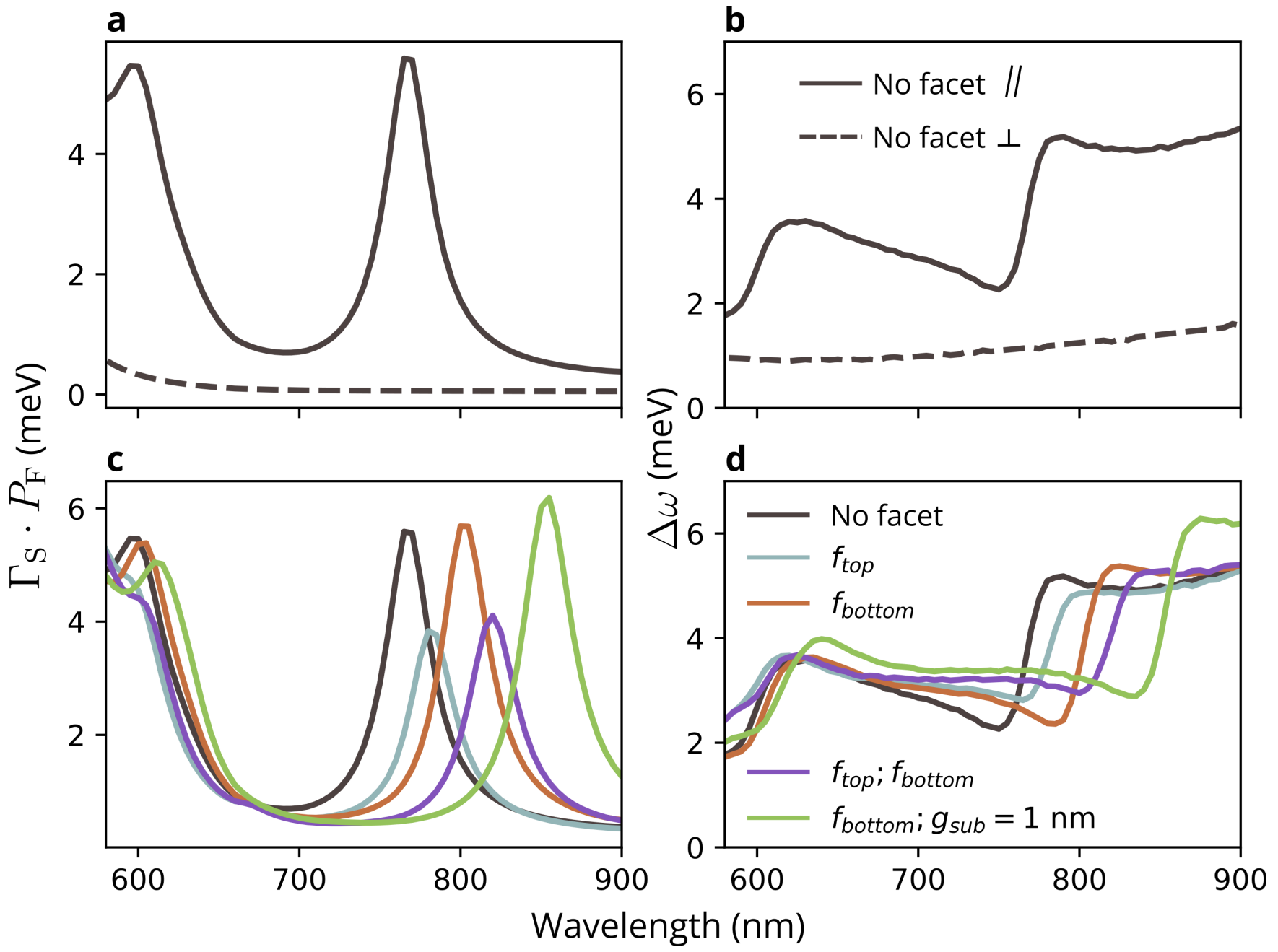

**FIG. S5.** Purcell Enhancement and Lamb-Shift in NDoM cavities. **a** Purcell broadening and **b** Lamb shift calculated for the NDoM spherical geometry wihout facets in Fig. S3a with a dipole placed at the center of the interparticle nanogap, i.e. 2 nm away from each Au nanoparticle (in reality the dipole is closer to one nanoparticle, see Fig. S1). The dipole is oriented either parallel (solid lines) or perpendicular (dashed lines) to the dimer axis. **c** Purcell broadening and **d** Lamb-shift calculated for the various NDoM geometries in Fig. S4 (same color code). The dipole is oriented parallel to the dimer axis and also placed at the center of the gap between the two nanoparticles. In all panels a spontaneous emission lifetime of the emitter in DNA $\Gamma_S = 1/3$ ns$^{-1}$ is assumed.

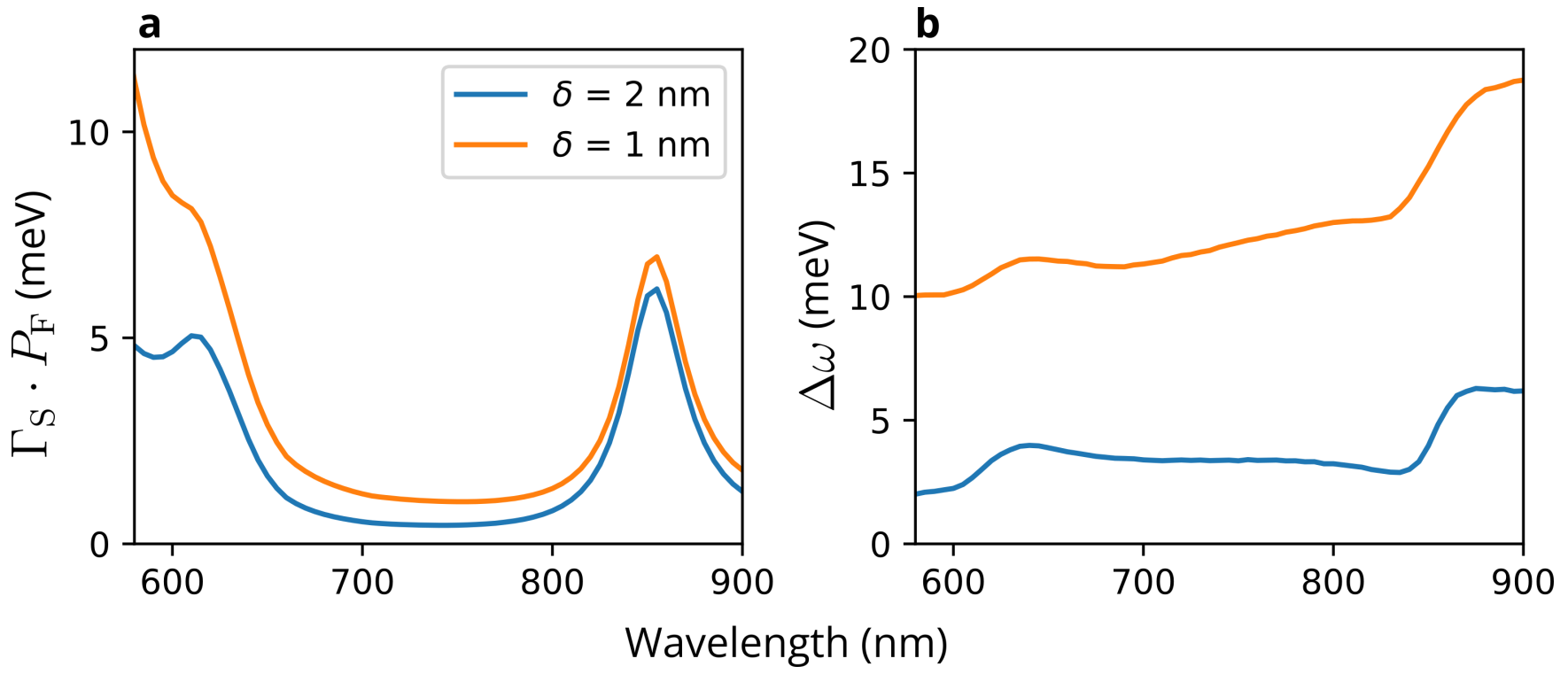

**FIG. S6.** Dependence of the Purcell Enhancement and Lamb-Shift on the position of the fluorophore. **a** Purcell broadening ($\Gamma_s \cdot P_F$) and **b** Lamb shift ($\Delta\omega$) spectra for a dipole situated in the gap between the two nanoparticles of a NDoM, calculated for two different distances (orange) $\delta$ = 1 nm and (blue) $\delta$ = 2 nm between the emitter and the surface of one of the Au nanoparticles. The geometry of the NDoM is depicted in Fig. S4c and presents flat facets in the gap between the Au nanoparticles and the Au substrate. The dipole is always located on, and oriented along the major axis of the dimer, and a spontaneous emission lifetime of the emitter in DNA $\Gamma_s = 1/3$ ns$^{-1}$ is assumed. The data in these simulations is used in the inset in Fig. 4b.

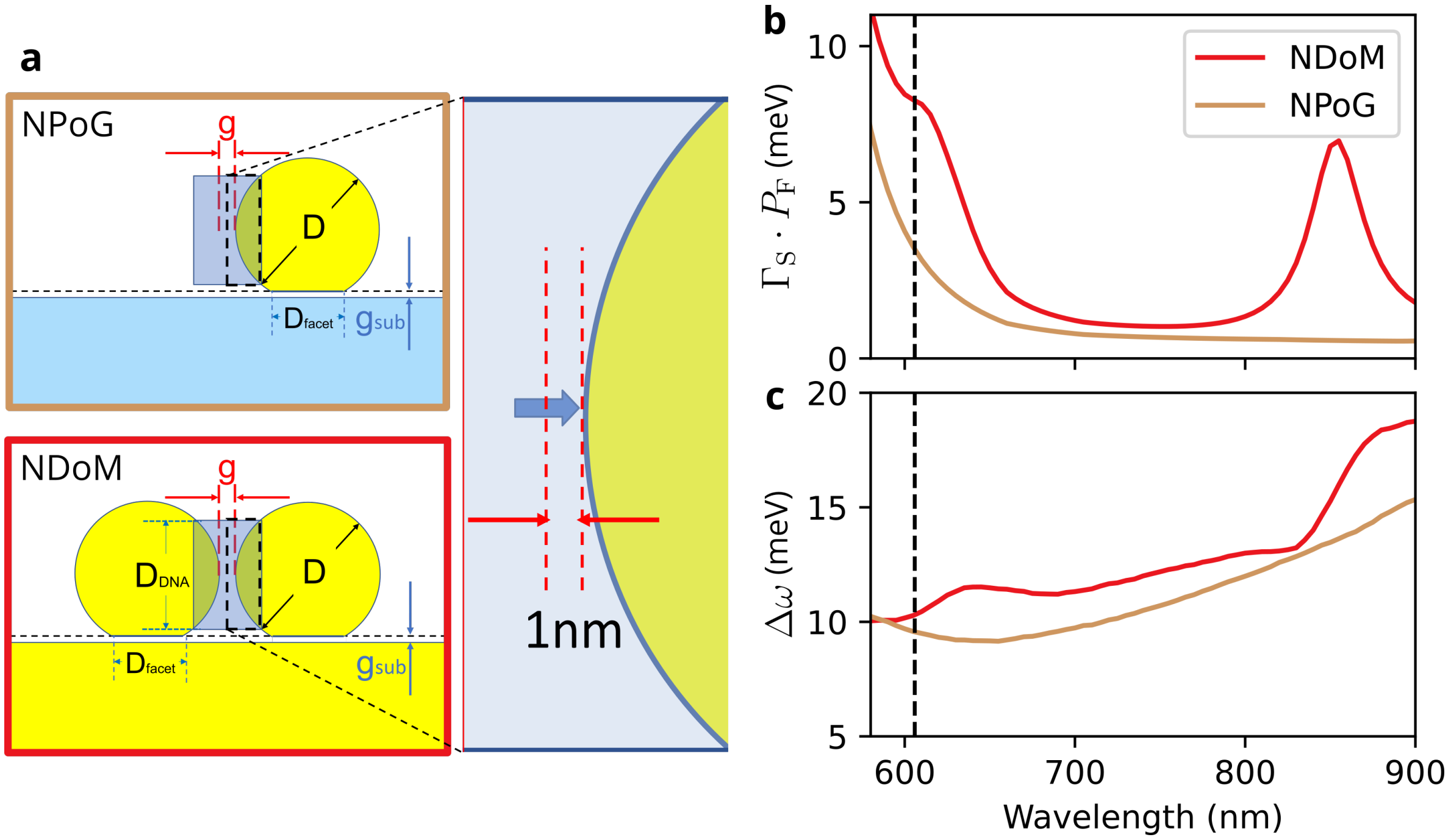

**FIG. S7**. Purcell broadening and Lamb-Shift in NPoG vs. NDoM : **a** Schematic representations of NPoG and NDoM geometries used. The dimensions for nanoparticles and facets are detailed in S3. As shown in **a**, the dipolar source is placed 1 nm away from the Au surface, and is oriented along the major axis of the dimer. **b** Purcell broadening and c Lamb-shift calculated for the two geometries. The vertical dashed line in **b, c** indicates the average measured ZPL frequency of the ATTO 590 fluorophore in DNA origami on glass at 4 K. A spontaneous emission lifetime of the emitter in DNA $\Gamma_s = 1/3\ \mathrm{ns}^{-1}$ is assumed.

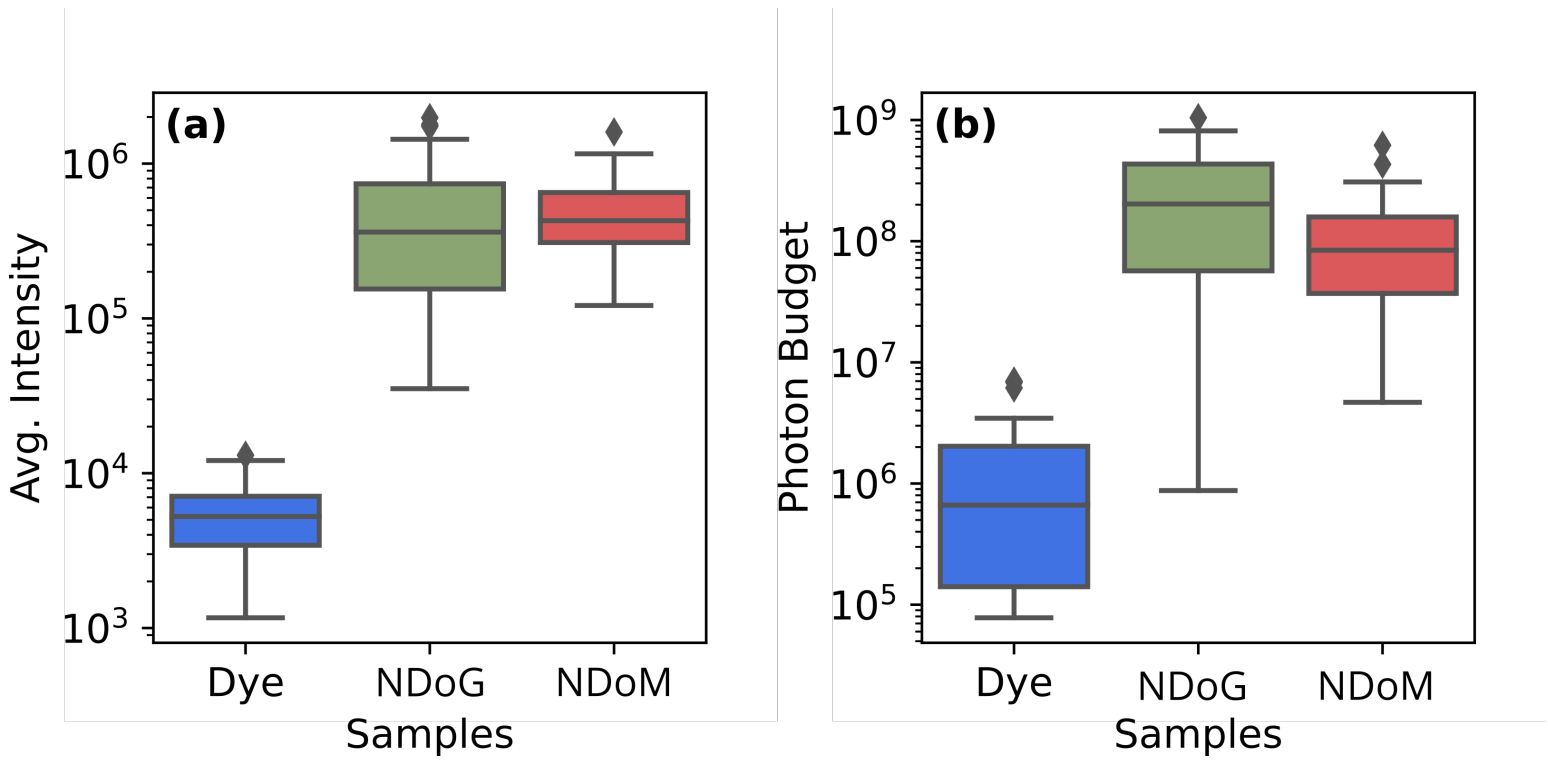

**FIG. S8**. Measured fluorescence enhancement: Average intensity enhancement **a** and total photon budget enhancement **b**. From each measured time-series, the ’ON’ spectra are extracted by subtracting the background collected after the fluorophore has bleached. Summation of counts from the average ’ON’ spectra gives **a**, while summation of all pixel intensities for the entire ’ON’ duration gives **b**.

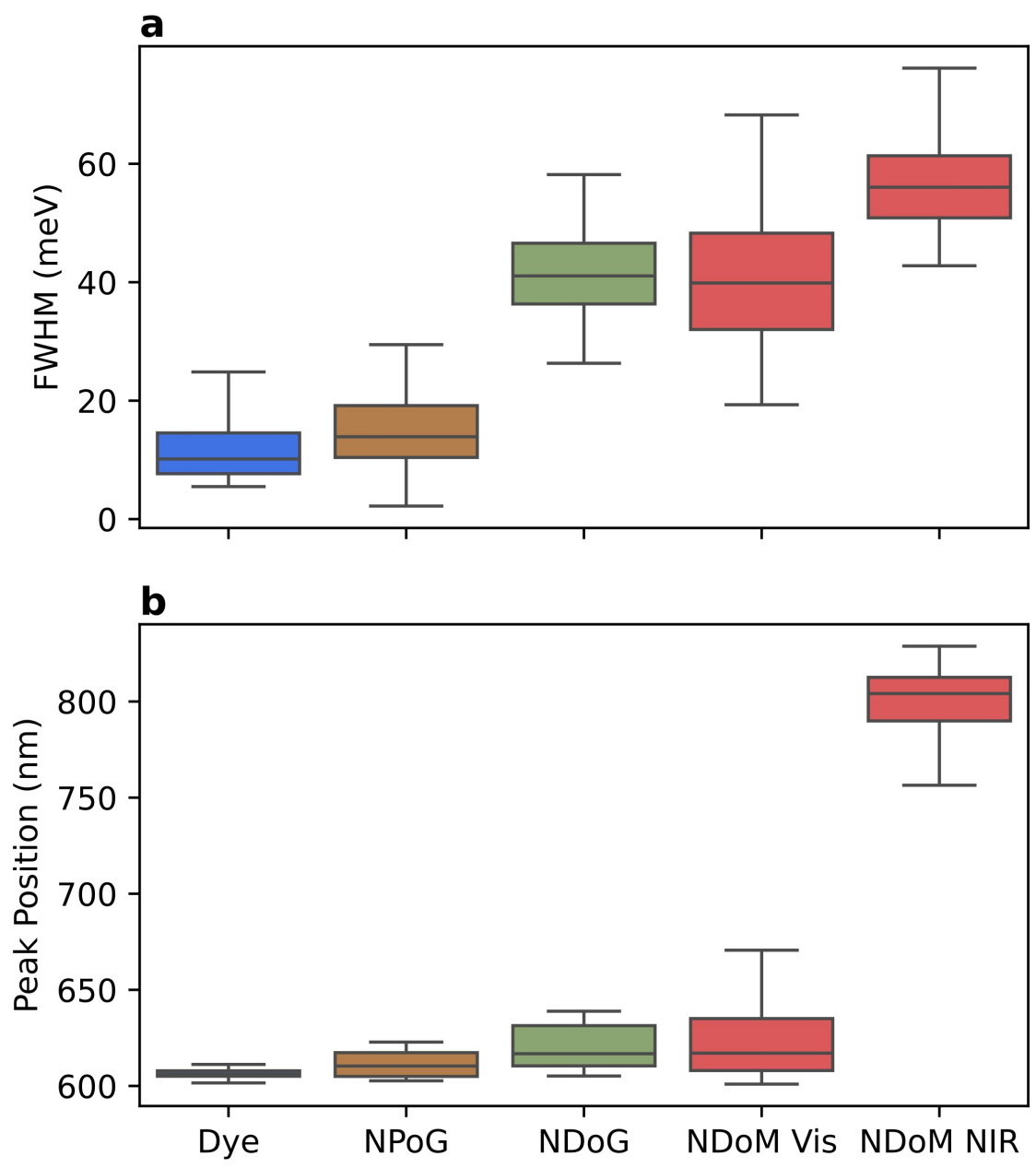

**FIG. S9**. Marginals of the data presented in Fig. 2, namely boxplots showing the distribution of FWHM, **a**, and position **b** of the peaks.

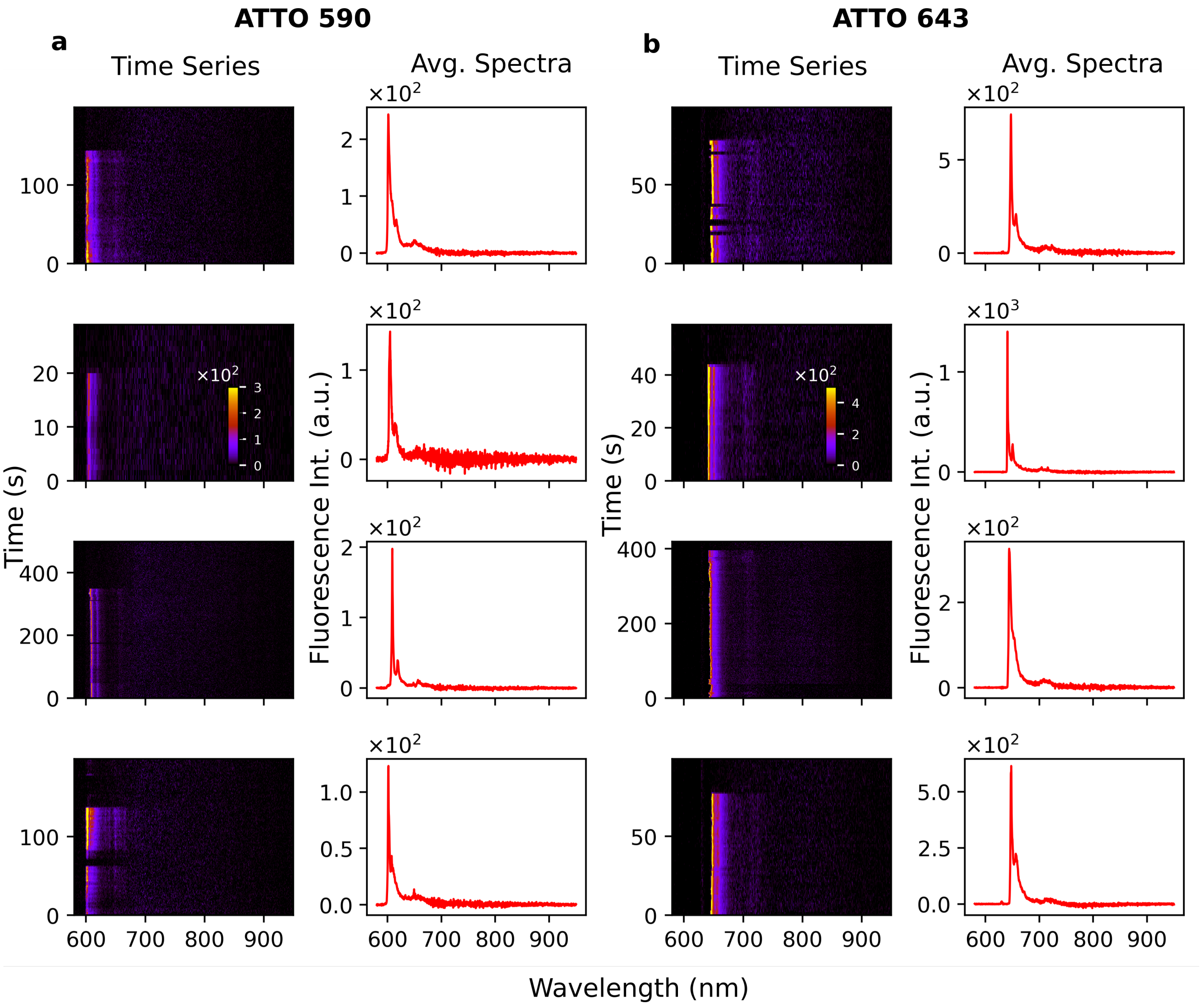

**FIG. S10**. Fluorophore spectra: Example time-series and average ’ON’ spectra for single fluorophores, ATTO 590 in **a** and ATTO 643 in **b**, measured at 4 K.

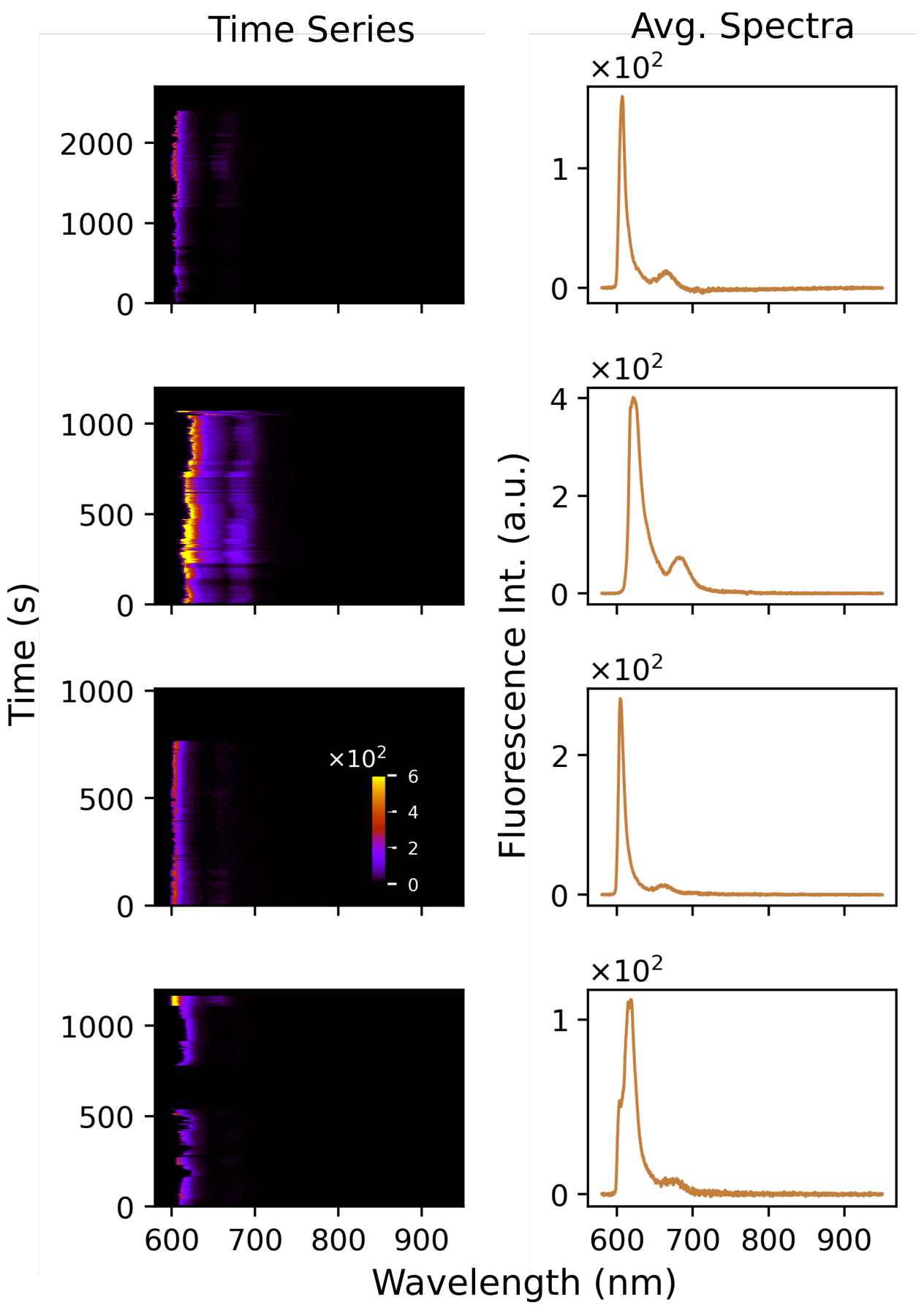

**FIG. S11**. Nano-particle on glass spectra: Example time-series and average 'ON' spectra from a few NPoG containing ATTO 590; measurements at 4 K.

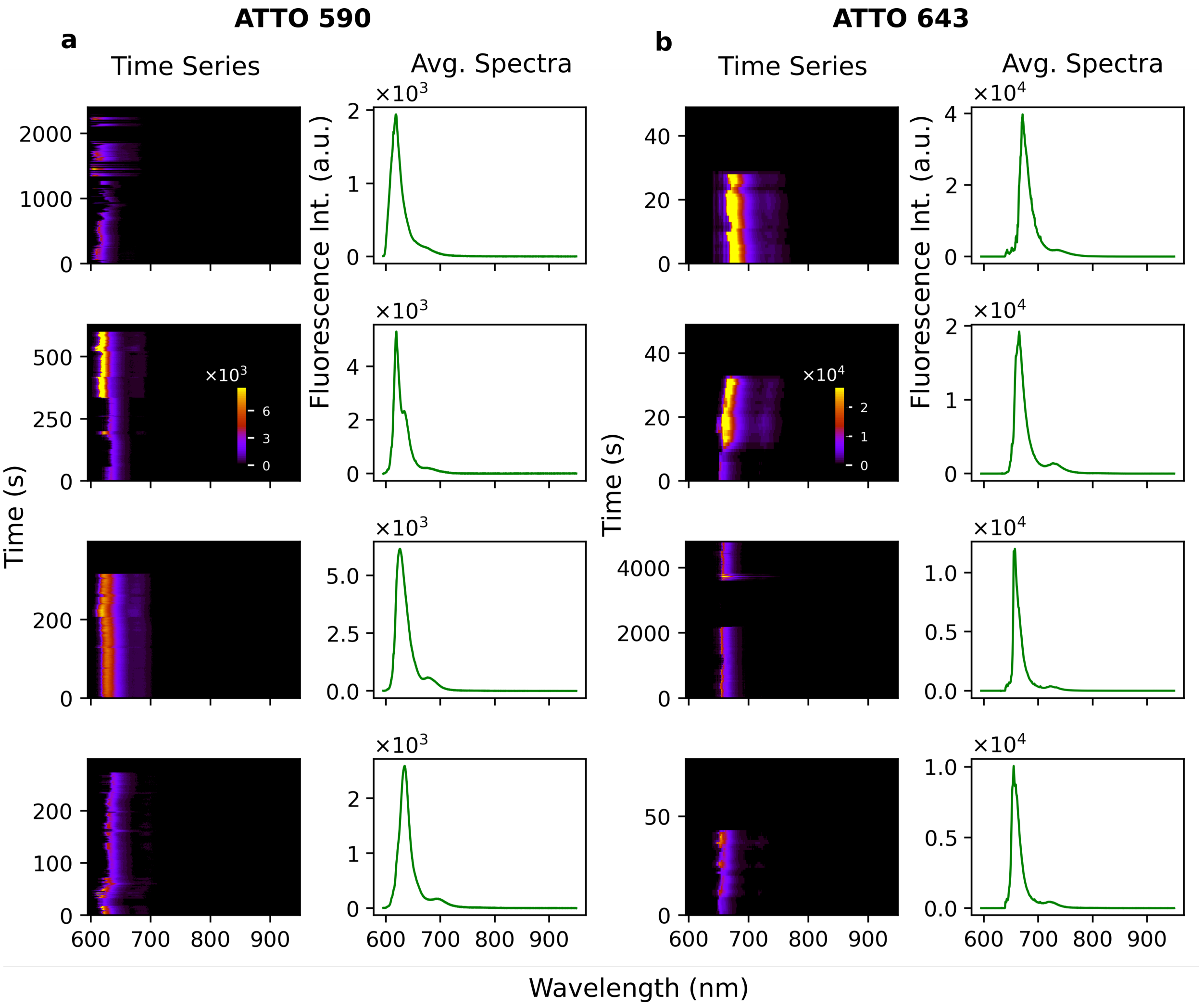

**FIG. S12**. Nano-dimer on glass spectra: Example time-series and average 'ON' spectra of single NDoG containing fluorophores at 4 K. ATTO 590 examples are shown in **a** and ATTO 643 in **b**.

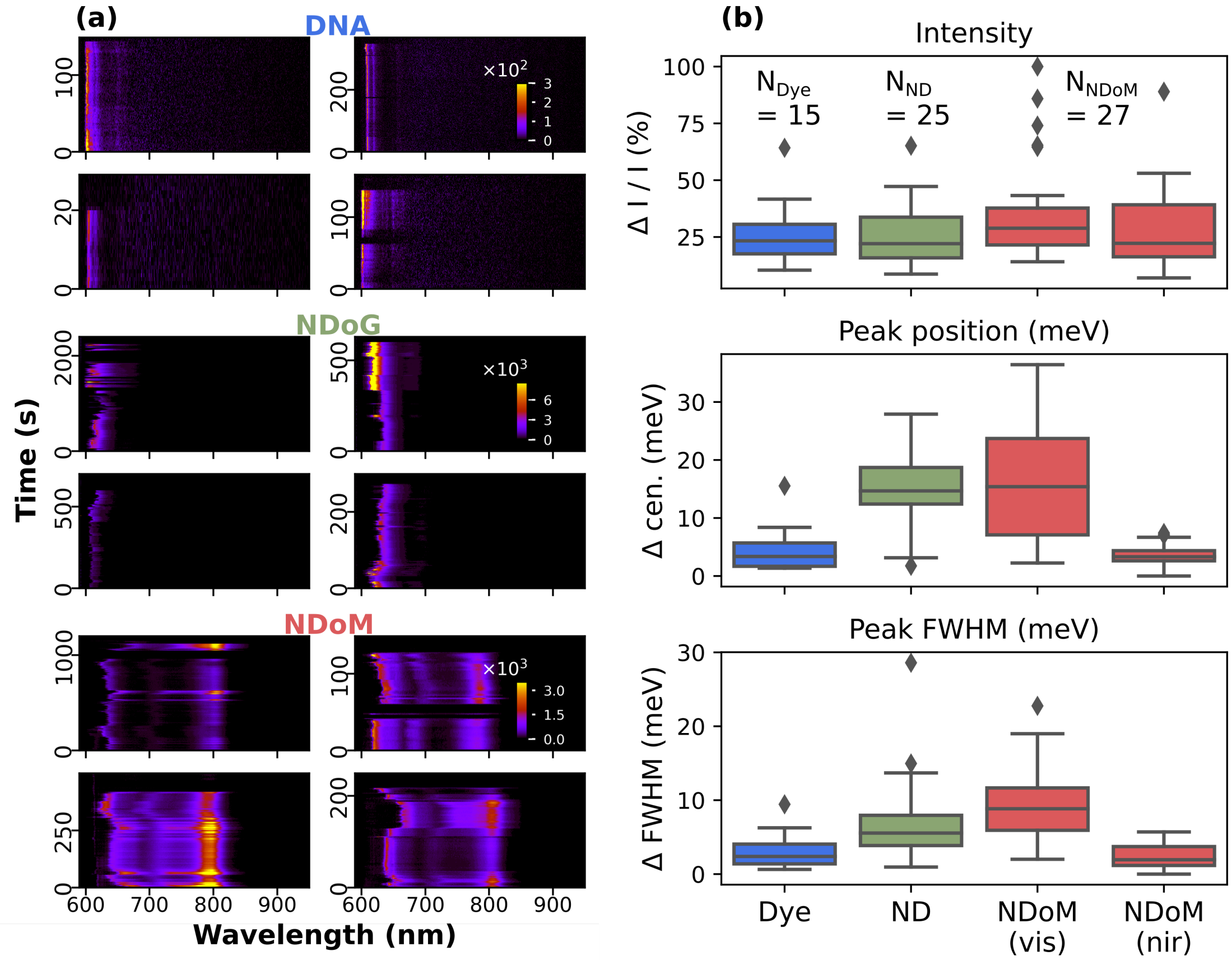

**FIG. S13**. **a** Full emission time-series of the nanocavities shown in Fig. 3 Excitation wavelength and power were 590 nm and 10$\mu$W, respectively. **b** Statistics of fluctuations of integrated intensity, peak position and linewidth for all measured structures (set sizes: $N_{\mathrm{Dye}}$ = 15,$N_{\mathrm{NDoG}}$ = 25 and $N_{\mathrm{NDoM}}$ = 27). Δ indicates the standard-deviation of the respective quantities, calculated over each time series. For NDoM spectra (red symbols) we analyse the ZPL emission (labeled 'vis') as well as the far detuned emission in the NIR region (labeled 'nir'). The central lines in each boxplot correspond to the median and the box shows the quartiles with whiskers extending to the rest of the distribution. Spectra classified as outliers are plotted as separate points.

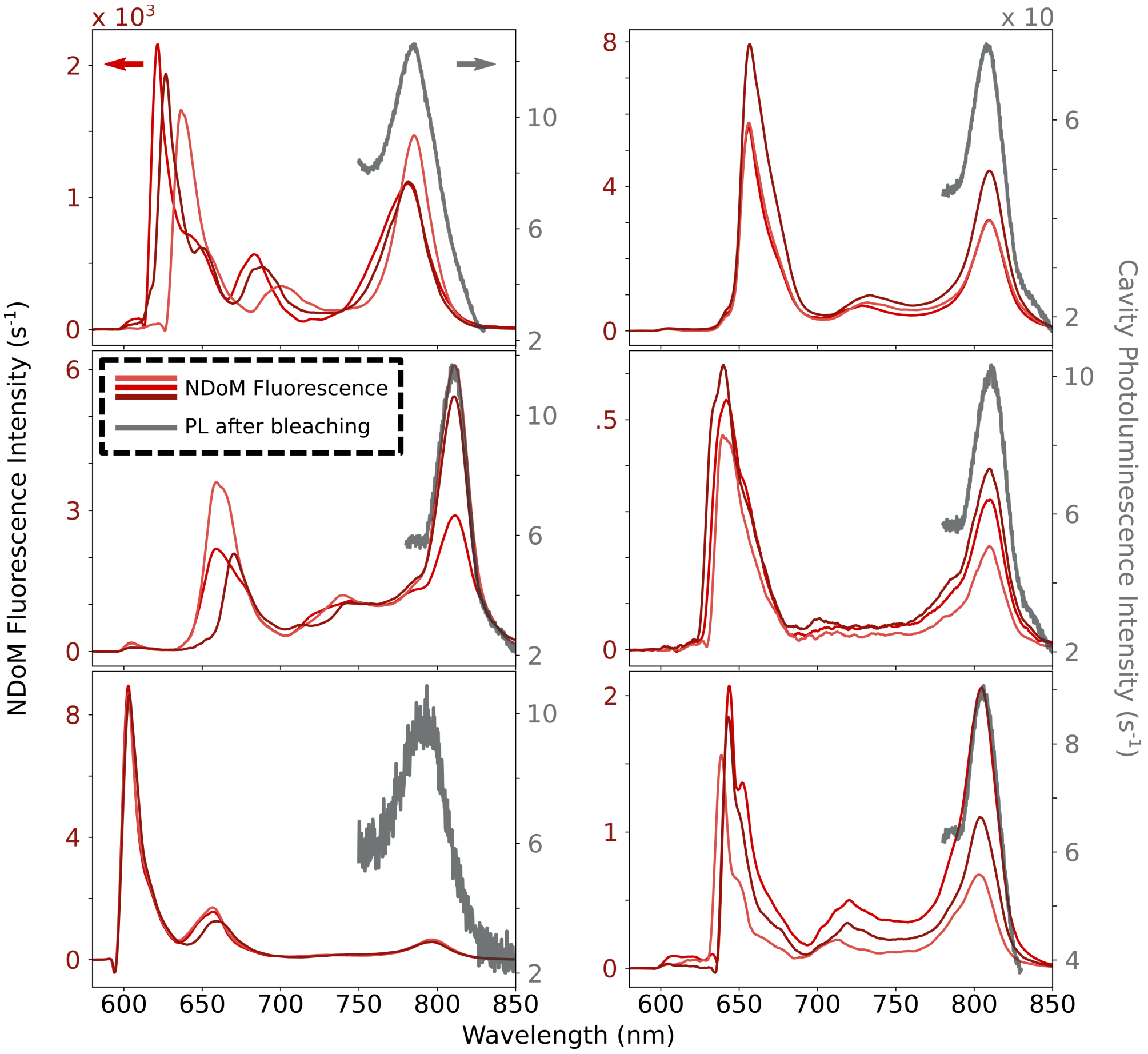

**FIG. S14**. Dye fluorescence and metal photoluminescence. Selected spectra showing significant NIR emission from various nanoparticles are shown with red curves. The red curves in each panel are extracted from the same time-series from a single NDoM. Residual, metal-induced nanocavity photoluminescence (PL) is also extracted from the same time-series after the fluorophore has bleached, i.e., when no ZPL emission is seen. This emission is typically 100 to 1000 times weaker than fluorescence and is plotted on a secondary *y*-axis. Excellent agreement is seen between the NIR fluorophore emission and the residual nanocavity PL peak, supporting our argument that the NIR emission arises through the plasmonic mode.

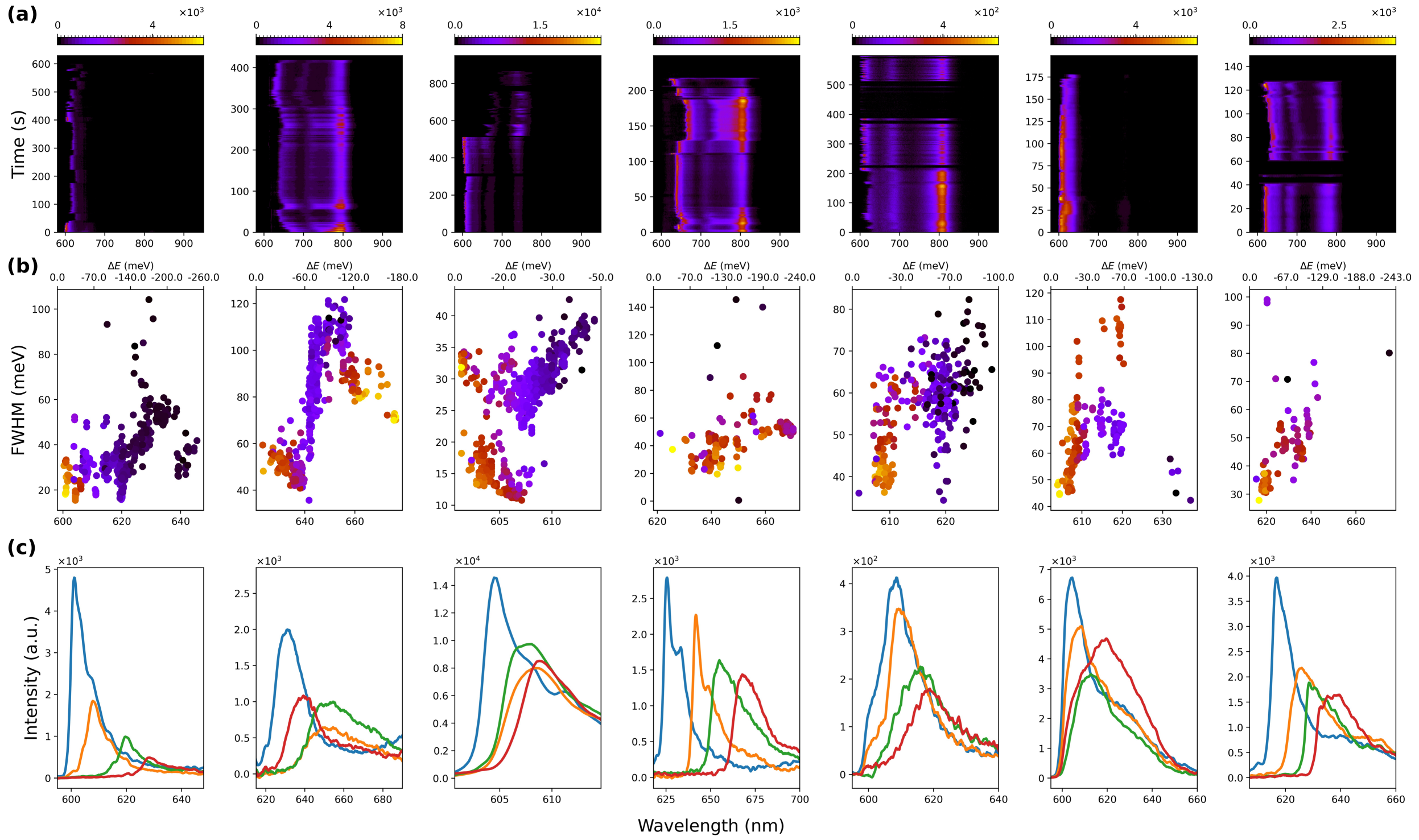

**FIG. S15**. Experimental signatures of Lamb shift. **a** Experimental time-series of ATTO 590 in NDoM at 4 K. **b** Correlation plots between the ZPL position and ZPL FWHM, with the color corresponding to the ZPL intensity. **c** Example spectra showing the general trend : a redshift of the ZPL is usually accompanied by broadening (increased FWHM) and decreasing intensity, consistent with an increase in total LDOS (mainly due to the nonradiative component).

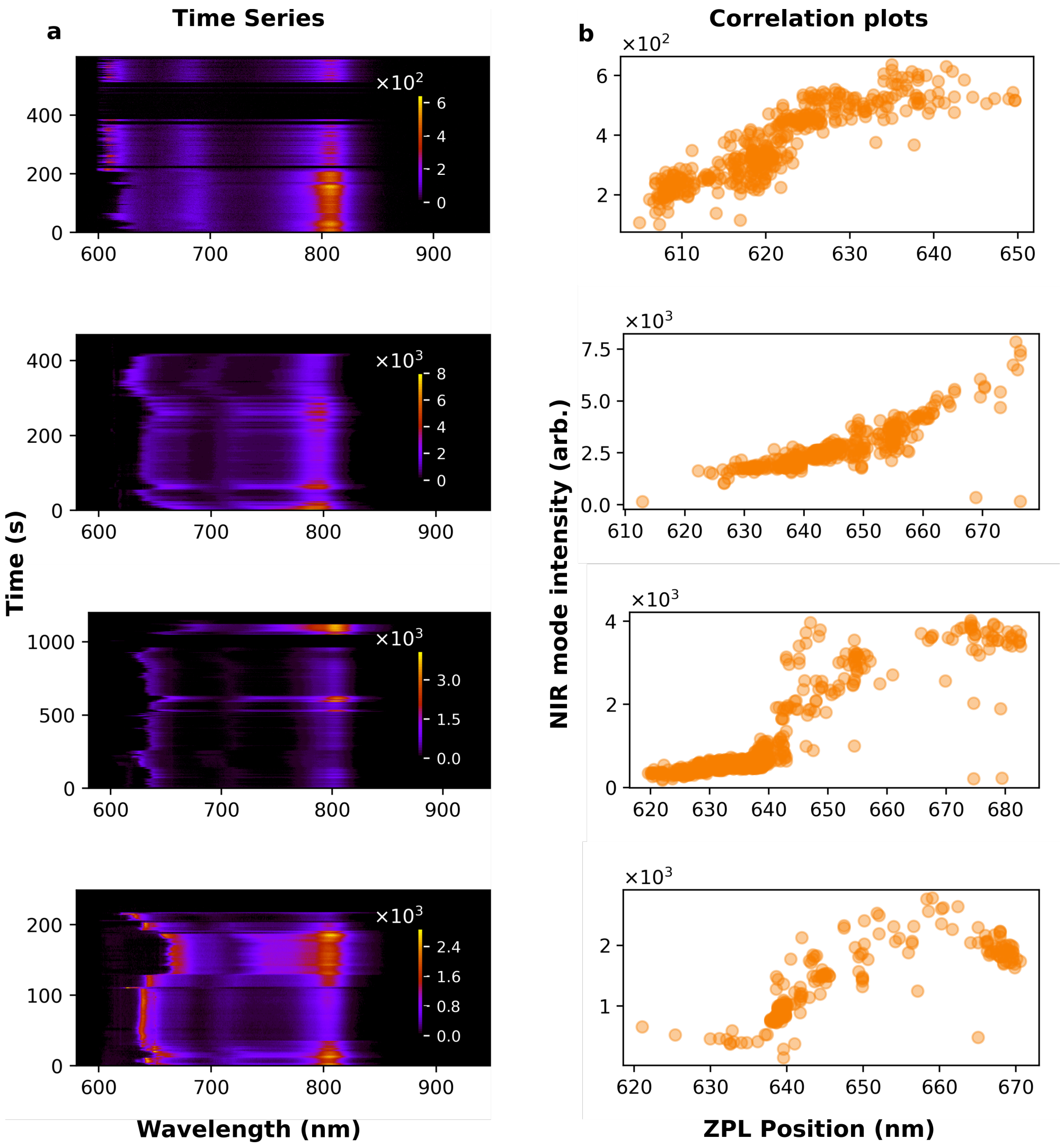

**FIG. S16**. Consequence of Lamb shift on NIR intensity. Time-series of ATTO 590 in NDoM at 4 K shown in **a** are analysed frame by frame. The position of the ZPL as well as the NIR mode intensity are extracted from each frame (1 s exposure), and the correlation plots are shown in **b**. A clear correlation is seen. This is consistent with our proposed model since a red-shifted ZPL would have a higher coupling to the NIR plasmonic mode, resulting in a larger intensity.

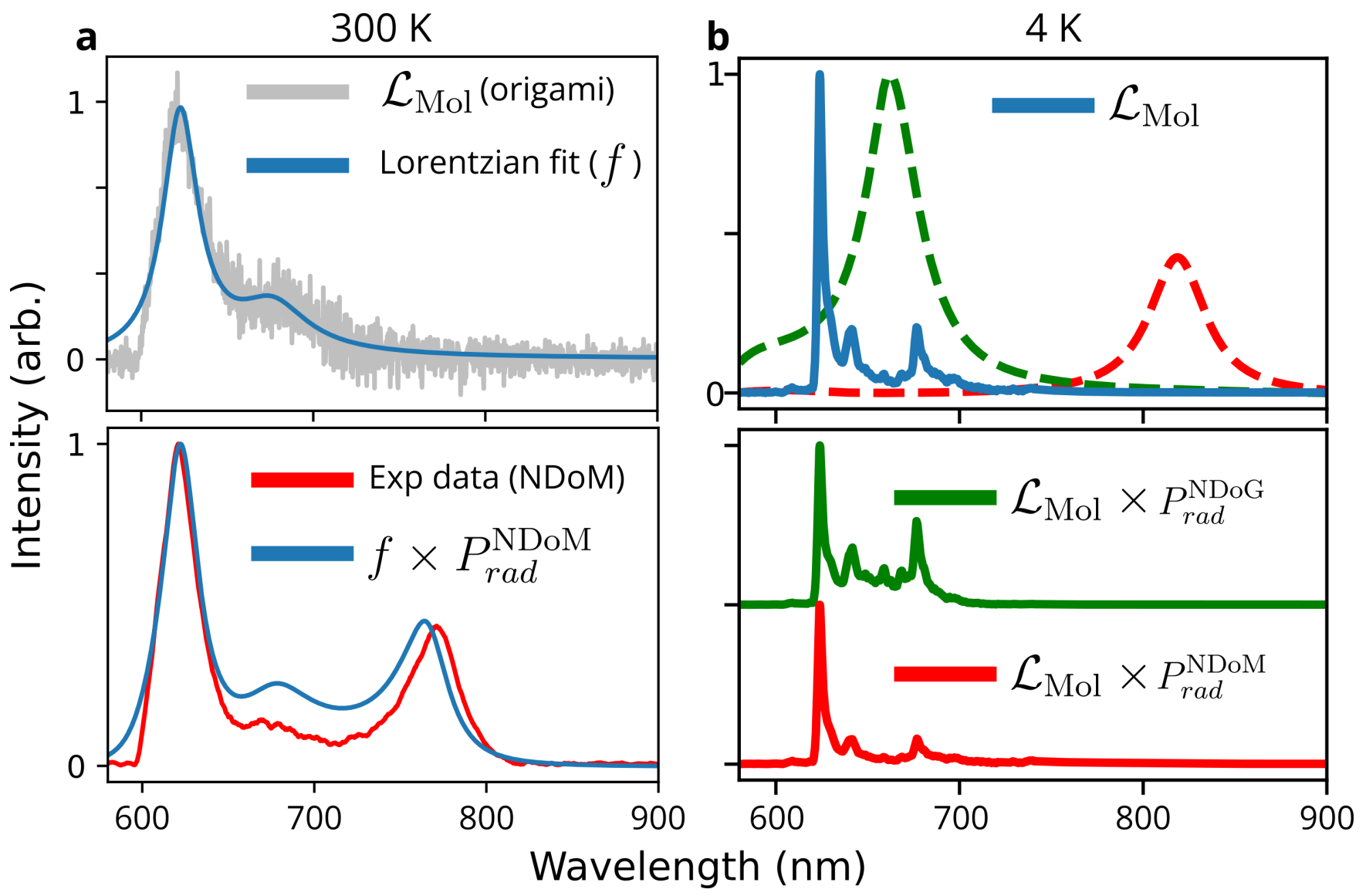

**FIG. S17**. Modelling spectral reshaping: multiplication model. **a** (top) Emission spectrum from a single-fluorophore at room temperature (grey solid line), along with a multi-Lorentzian fit (blue solid line). (bottom) Reshaped spectra obtained via multiplication, and compared to experimental data at room temperature. **b** (top) Emission spectrum from a single NPoG at 4 K (blue solid line), along with the rescaled radiative LDOS for NDoG (green dashed line) and NDoM (red dashed line). (bottom) The reshaped spectra naively obtained by multiplication according to eq. (1) are shown in corresponding colors for NDoG and NDoM respectively, none of which agrees with the experimental observations.

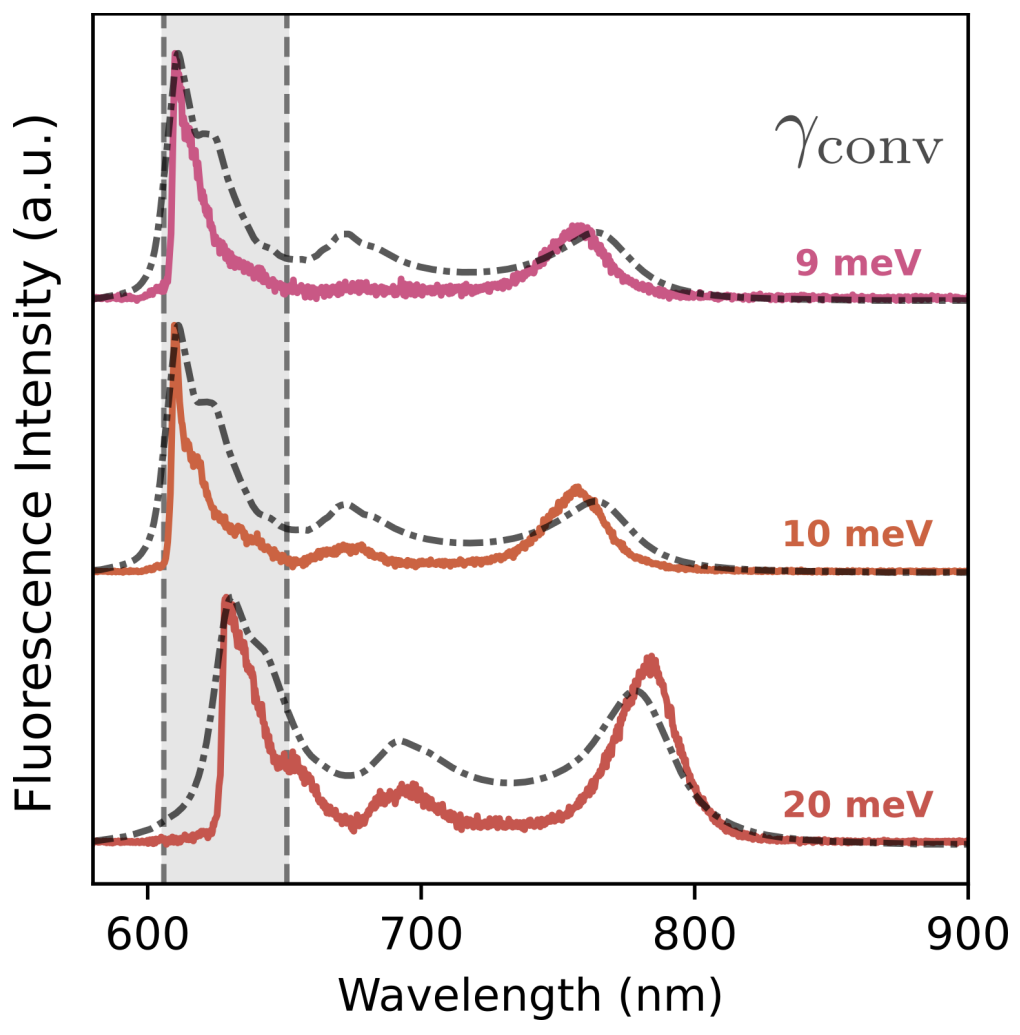

**FIG. S18**. Convolution model: A few more examples similar to Fig. 4c.

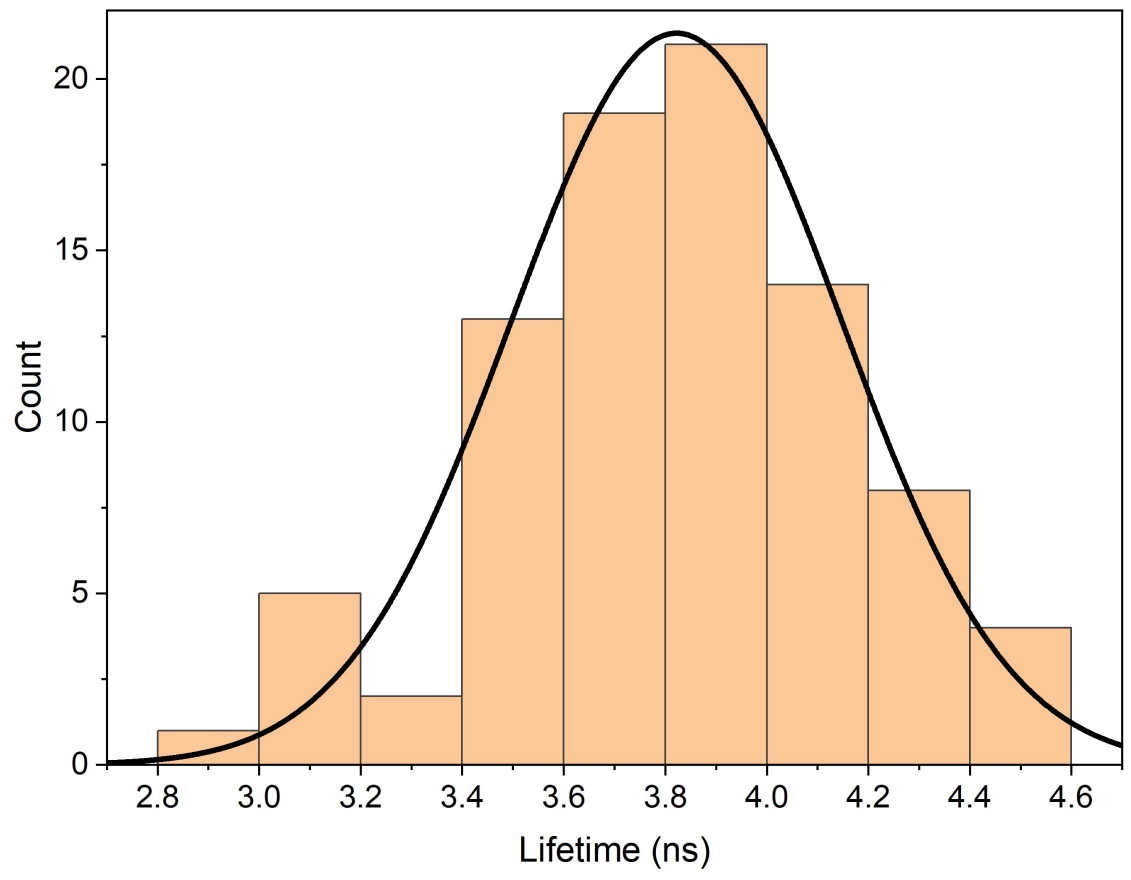

**FIG. S19**. Fluorophore lifetime measurement. DNA origami with ATTO590 on glass in air is measured at ambient conditions. The lifetime histogram of all measured points is plotted, and fit to a Gaussian, giving a mean lifetime of 3.8 ns. Since the quantum yield is characterised by the manufacturer to be 80%, the radiative lifetime is estimated as 4.7 ns

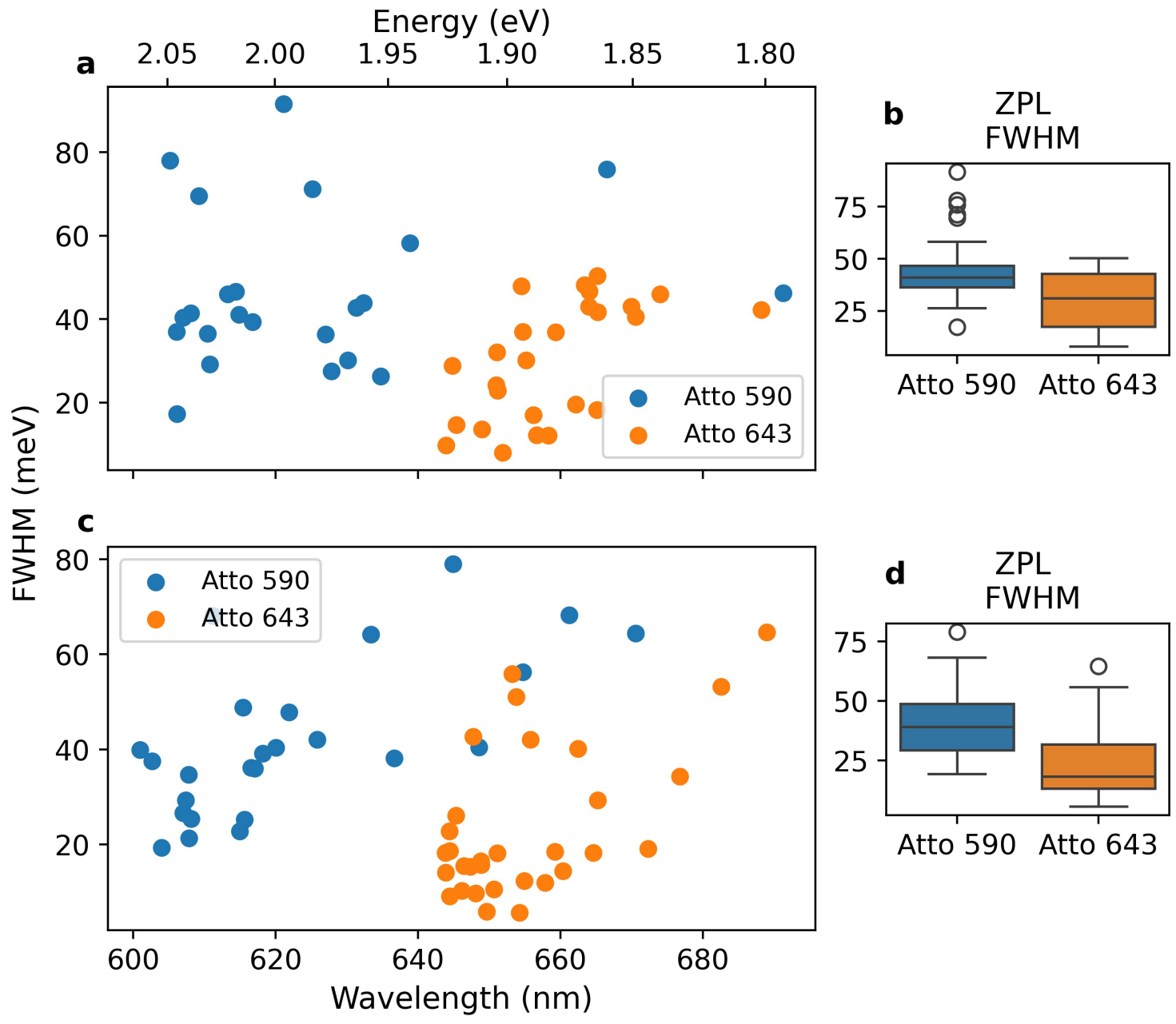

**FIG. S20**. Comparison between ATTO 590 and ATTO 643. Scatterplots of ZPL positions vs. ZPL linewidths for fluorophores in NDoG **a** and NDoM nanocavitites **b**. The linewidths are also compared in the boxplots on the right.

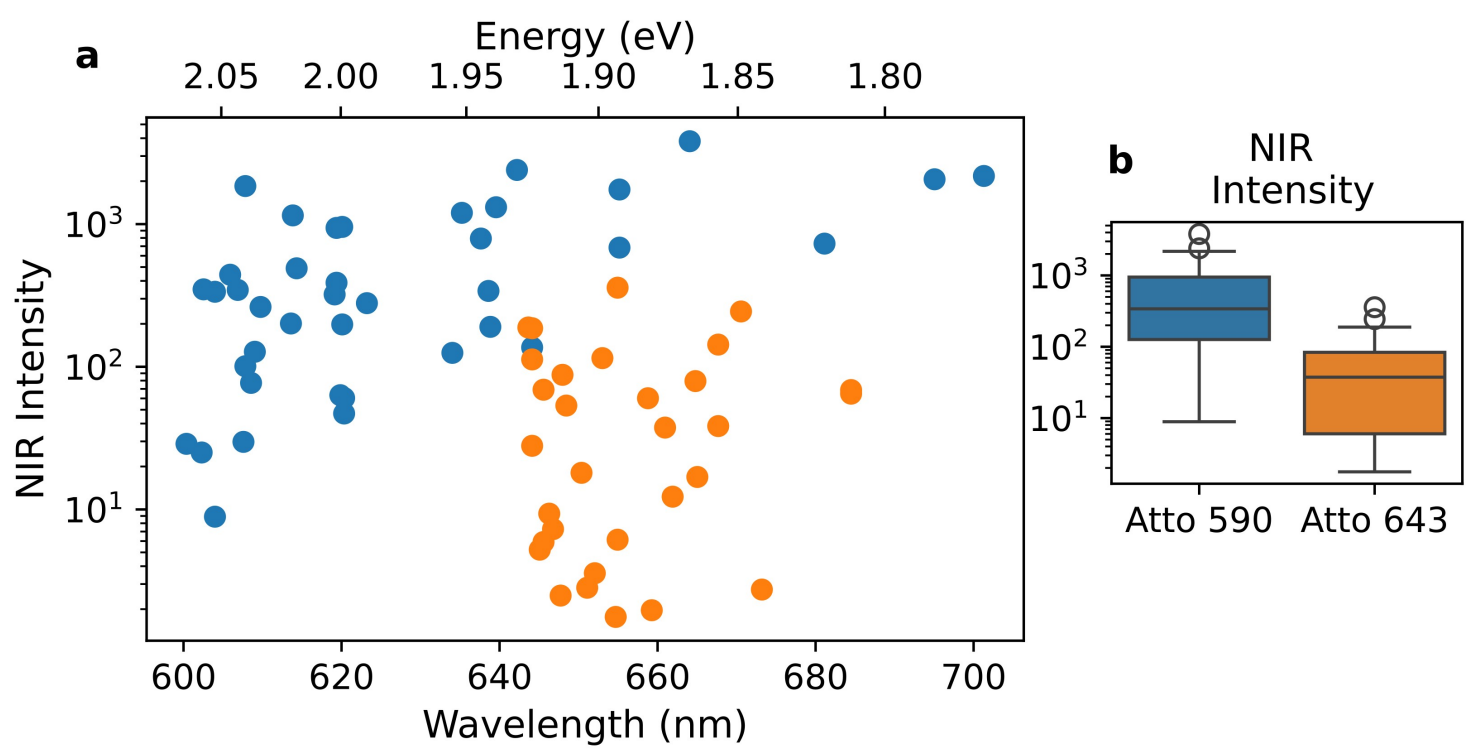

**FIG. S21**. Comparison between ATTO 590 and ATTO 643. Scatterplot of ZPL position vs. integrated NIR intensity, which is also analyzed separately in the boxplot

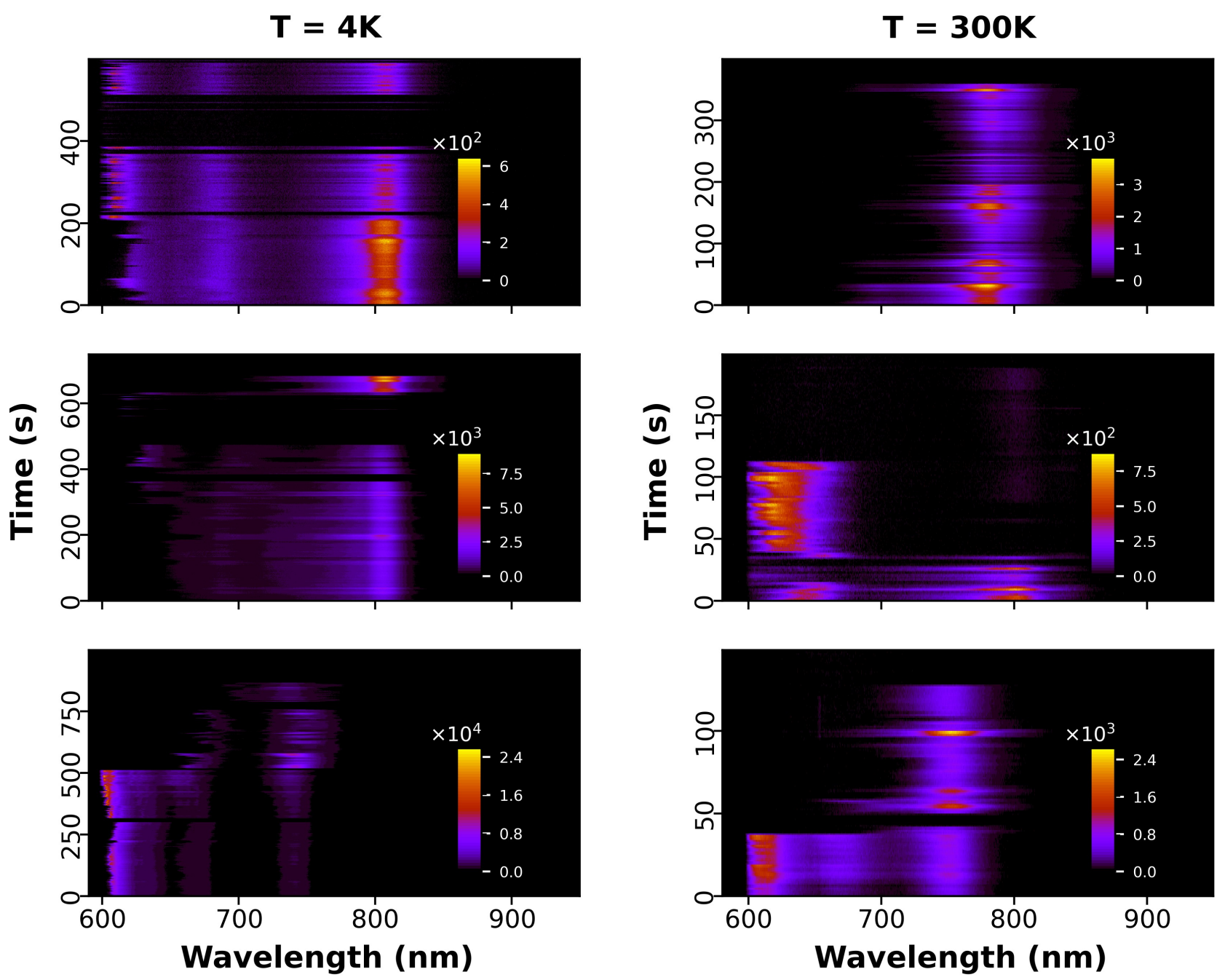

**FIG. S22**. Extreme reshaping events. Fluorescence time series from individual ATTO 590 fluorophores in NDoM at $T$ = 4 K (left column) and $T$ = 300 K (right column) showing extreme fluctuations in the ratio of NIR to ZPL emission. Such events seem to go beyond the level of approximation used in our model.

TABLE S1: List of DNA origami staples

| Staple type | Sequence |
|---|---|
| Core Staple | CCAAGTACCCATATTTCGACGACAATCATAAT |
| Core Staple | ACCAGAACCACCACCAGTTCCAGTAGTGTACT |
| Core Staple | CGCGTACTGGTAATATGAGTAAAAGACCTGAA |
| Core Staple | CCAACGCTCTACAATTCGTAGGAAAAGCAAGC |
| Core Staple | AGACAGCCCGATAGTGGAGCCTTAAACGGG |
| Core Staple | GACTGTAGTTTTGATGCGGGGTTCATTTGG |
| Core Staple | TTTAACGTAATGGAAACTATTAATGATAGCTT |
| Core Staple | AGTAATTCCATCCTAAAAGAACGGTATAGAAG |
| Core Staple | GGTCGCTGCGCATTAAAGCAGATAATTGCG |
| Core Staple | AGCACTAAGCCCGAACCCACCAGAGGTTAGAA |
| Core Staple | AAACACCATCAGGACGTTGAGATTGTAAAATG |
| Core Staple | AGATTAAACGCTCATTTAGTAAAGGTAA |
| Core Staple | ATAATAACTAGCAATAGTCAGAGGAATGAAAA |
| Core Staple | AGAAGGATCGGATAAGAGCAAGCCTTCGTCAC |
| Core Staple | CAAGCGCGGTAATGCCCACCCTCACAATGACA |
| Core Staple | CAAAGTTAAGAAAAGTGACGGGAGCATAAAAA |
| Core Staple | CGCCTCCCTCAGAGCCAGCGTTTGGTAATCAG |
| Core Staple | AATTTAGTCATTCCTTTACGAGGAGGTTT |
| Core Staple | GAATTAGACGTCACCGTATTTTGTGCAAAGAC |
| Core Staple | CCTCATTTGTTTTAACGCTGAGACGCCAGCAT |
| Core Staple | TACCGAGCGCCAGGGTCGCCATTCCGACGACA |
| Core Staple | TGCGCTCAAAAAGAATCCGCCTGGCCCTGAGA |
| Core Staple | TACTAGAACCTCCGGCAACATAGCTAATTTTC |
| Core Staple | CCACATTCAGGCTGGCAGTAAATTGATTATAC |
| Core Staple | GTACCAGGTAGGATTAGATACAGGAAGCGTCA |
| Core Staple | GCACCCAGAACGAGCGGAGAATAAAATTAACT |
| Core Staple | TGGTCAGAAAGAAAGTTATTATTTCAGG |
| Core Staple | AATATATTTCATCTTCGGGTAATATAGGAATA |
| Core Staple | AATAATAAAAGGAACACACTGAGTCAATAGGA |
| Core Staple | GATATAAGATTAAGAGGGGGTCAGGAAAGCG |
| Core Staple | TGATAATCCTCAGGAATAACAACCTCACGACG |
| Core Staple | GGCCAACATAAAACATCGGTCAGTTCAATATC |
| Core Staple | AAATGGAGGTGAGGCGCCATTACTTTAGG |
| Core Staple | GCAAACAATTAAGCAATATTTTAATACAAAAT |
| Core Staple | GGTACGCCAGGCCACCCCAGAACACGCTCATG |
| Core Staple | GAGTAATTTTAGTAATGACCATATCTGCGAA |
| Core Staple | TTGTAAAAGGTCGACTTAAAGCCTAATTGCGT |
| Core Staple | AGCTGATTTTTTCACCCTCACATTGGGGTGCC |
| Core Staple | TTAGATACCAGTTGATAAATATGCCAAAGCGG |
| Core Staple | GCTTAGAGCTACTAATTACCAAGTATGCAATG |
| Core Staple | AGCGTAAGCTATTAGTACGCTGAGACCTTGCT |
| Core Staple | ACCAATAGGGAACAAATGAGGGGAAGGCTGCG |
| Core Staple | CCTACCAATCGTCGCAGTACATAGACTAC |
| Core Staple | ACCCATGTATAAGTTACTTGAGCTTGCTCA |
| Core Staple | TGACAGGAGGTTGAGGCCAGAATGTGCCTTGA |
| Core Staple | CTTGATACCTCATAGTCTGTATGTGTATCA |
| Core Staple | GGTAATAATCAGGGATTGCCGTCGGCGGAGTG |
| Core Staple | ATTGCATCGGATAGCAAACAGTTGAAAAATC |
| Core Staple | GAACCTCATCATATTCTCGACAACAACAGTAC |
| Core Staple | AATGGGATTTGTTAAATTTAAATTATCTACAA |
| Core Staple | GGCAATTCATCATACGTAAAGATAAGTATT |
| Core Staple | TACGTTAATGAATAAGCCGGATATACCTGCTC |
| Core Staple | GTGCCGTACCGATTTACTGCGCGTCTACAGGG |
| Core Staple | GCTAAACACTCCAAAATGCGCCGAGCAGCGAA |
| Core Staple | ATAGAAAAGAATCAAGAATCACCAGTAGCACC |
| Core Staple | TGAAAGAGTTTTCATTTATCCTGTACACTA |
| Core Staple | CCTAAAACATCTTTGACCGAACTGTACAGACC |
| Core Staple | ACCACGGAACCGTAAACTAAAGGAGCCGAA |
| Core Staple | AGGCGCATAACTAATGTTAAGAACTGGCTCAT |
| Core Staple | GCCTTGCTATGGTTGCGATTTTAGCGGGGAAA |
| Core Staple | GATAGCCCGAGATAGAGACGCTCAAACTATCG |
| Core Staple | AATCCTTTCAACTAATACCCTCAAATTAACAC |
| Core Staple | CCTGAGTAAGGCCGGAATGAACGGACCCCGGT |
| Core Staple | AAGTGTAGGAACGTGGGTTTTTTGGGGTCGAG |
| Core Staple | GCGCTAACAAACGTCAAAAGAAGGGAAGGT |
| Core Staple | ATCAGGTCAATGCTTTGTCCAATATTACAGGT |
| Core Staple | TAGCAAAAGAGAATCGGACAGTCAAATTCGCG |
| Core Staple | GAGTGTTGTTCCAGTAATCGGAAGGGCAAC |
| Core Staple | TGCCTGCACGACGGCCTTCTGGTGTCCAGCCA |
| Core Staple | GTGCATCTGCAAATATCAGCTCAGCTGATA |
| Core Staple | ATCAACAAGCTAATGCCATAGTAAGAGCAACA |
| Core Staple | TCTCCGTGGAACGCCACCCAAAAAAGTCTGGA |
| Core Staple | AGGCTATCATCGGTTACGCAAGGTTTGACCA |
| Core Staple | TCATAGCCCCCTTATTGCCACCCTCAGAACCG |
| Core Staple | ATCGGTTTATGAATTTTTAGCGTAAACCGCCA |
| Core Staple | AAACACTCGAAAGAGGCAGGGAGTATATATTC |
| Core Staple | CATGTTCATAGATAAGTCGAGAACTCATTACC |
| Core Staple | AATTAATGTTATTTCAGTACCAAATAGCTATA |
| Core Staple | GAAATACCACAGTGCCCTTTAATGGCCGTCAA |
| Core Staple | GCTTTCCGTTGCAACGCACAGACCATCAAC |
| Core Staple | ACAGGTCAACCCTCGTACACCGGAATAAACAA |
| Core Staple | GCCAGCCAGCACCGCAGTGCCAATTTTTAT |
| Core Staple | GGAGAGGCGCAAGCGGAATCGGCAAAATCCCT |
| Core Staple | GCTTATCTTGTTTATAAACAGCAAGAAA |
| Core Staple | CCACCCTCAGAGCCACTCATCGGCTAGCGTCA |
| Core Staple | AGTCTCTGAATTTACCGAGCCGCCTCCTCAAG |
| Core Staple | TATACCAGGAACGAGTTGACCTTCCGCAGACG |
| Core Staple | GCTTTTGCTAAGGGAACCCCCAGCGGGCTTGA |
| Core Staple | TACATGGCCGCGTTTCACCCTCAGAGCCGCC |
| Core Staple | TTGTATAAGCCAGTTCGGCGGAGATTAAGT |
| Core Staple | ATTACCATACGGAAATGTTTACCACATACATA |
| Core Staple | CAGGCAAAATAAAGTGCTAGAGGAGCCAGGGT |
| Core Staple | AAGGTGGCTGAACAAAGCTATCTTGAGCCTAA |
| Core Staple | GGTTTTTCGCCCTTCAAGCCCGAGATAGGGTT |
| Core Staple | AATTATCAAATATCAAAGATTAGACGCGAACT |
| Core Staple | GCCGGCGACAAATCAAACTCCAACGTCAAAGG |
| Core Staple | AATCAGTGAGAATCCTATGCGCCGAACCACCA |
| Core Staple | TAGATAATCAAACAATCTGATTATATTGTTTG |
| Core Staple | TTTTCATTATGTTTTTCCCAATAATCAAAA |
| Core Staple | TTTGCCAGTAAATCAAAAATCAGAGTATTAAA |
| Core Staple | CCTTTTTTCAGATGAAAATGGAAGAGGAGCGG |
| Core Staple | TATCATCGATGAGGAAGAGGGTAGTTAAACAG |
| Core Staple | CAATGAAAGGAATACCAGAAAATAGCGCCAAA |
| Core Staple | CCATCACCACGTGGCGGCGCTAGGGCACGTAT |
| Core Staple | CACCCGCCGAGCTAAAGTGAGACCCCTAAA |
| Core Staple | GAGTTGCAGGTTTGCGTTTCCAGTCATACGAG |
| Core Staple | GCGAAAAGGAGCGGAGAAAGGAGCTAAACA |
| Core Staple | AGAATAGATTTTTTCACATAACCGTAAAGGCC |
| Core Staple | TGGGTAACTCGAATTCCCACACAACGGGAAAC |
| Core Staple | CATGTTACTCGGAACGTTTCCATTTAATTGT |
| Core Staple | CCTTAGAATTCTGAATTATACAGTTCGTATTA |
| Core Staple | CGCCTGCATACATTTTACCCTTCTGAGTCTGT |
| Core Staple | CCGGAAGCGCGCCATTTTTCCCAGCGTCGGAT |
| Core Staple | TGAAGCCTTTACAAAAACGTCAAAGTAATTGA |
| Core Staple | GGAGGCCGCAAATTAAAGAACTCAATCGTCTG |
| Core Staple | CTATCATACGCACTCATCCTGAACGCTATTTT |
| Core Staple | CTATTAAACGGTCACGGAGCTTGAACAGGAAC |
| Core Staple | AGCTCAACTGGGGCGCGAGCTGACAGAGCAT |
| Core Staple | ATTAAATGTCCTGTAGTCATATGTTAATCGTA |
| Core Staple | ATTACCTTAATTTCAAGAACGGTGACCAACTT |
| Core Staple | TAGCAGCCTAGCAAGCGATTAGTTAAGAAAAA |
| Core Staple | GGGAGCCCAAGCACTATTGGAACAAGAGTCCA |
| Core Staple | TAATATCCTGTCCAGAAACAACGCGCTTAATT |
| Core Staple | AAAGCTAAAGGTCATCCGTTCTATTTTTTA |
| Core Staple | ACAACCATACTACAACCAGTTTCAAGAGGGTT |
| Core Staple | CCTTTTTACCAGAAGGAAAGAAACCACAATCA |
| Core Staple | TTGAATGGAATACGTGAGGAAAAAATATTACC |
| Core Staple | AAAAAAGGACTTTCAAGCCTGTAGGCCACCAC |
| Core Staple | GACAAAAGATCGATACCGGAAAACCGGAAC |
| Core Staple | TTTTAGAAATATTCAATGCCTGAGCAGGAAGA |
| Core Staple | TATAAATCCTGCCCGCTATTGGGCTCCCCGGG |
| Core Staple | TAATGAGTGCGCTTAGAGAAGTGGCTTGCA |
| Core Staple | CGTTTTTAGACAGATCTTTAATCCCTGTTT |
| Core Staple | GTATCGGCAGAAAAGCTCAAAAATAATCACCA |
| Core Staple | GCAAATGAATGTCAACCAGCTTTAATATTT |
| Core Staple | AGAAAGATACAAGAAGCTTGCCGAGATTTG |
| Core Staple | GCGCCCAATTTACAGATCTTTCCAACCGAAGC |
| Core Staple | CCATCACGATTAAAGGTTTGACGAGCGCTGGC |
| Core Staple | AGACTTTAACATTTGAAAAGCATCAGCCAGCA |
| Core Staple | TAATCTTGTCATCAGTTGGGAACAGAAAAC |
| Core Staple | CGAATTATGGCGAGTAGATTTAGATAAAAAT |
| Core Staple | ATTATCACGCCAGCAATTTGCCTTATTTTCGG |
| Core Staple | TCAATATGCCCTCATATAAAGCCTAAAGGTGG |
| Core Staple | AAACTAGCAAAATCTGGATTTAGTCAAAAG |
| Core Staple | CAGGGAAGAGGCTTGCAAAAGAAAATCTTA |
| Core Staple | CTTTTTAAAAAGCCTGACAGTAGGCAACATGT |
| Core Staple | CAGTACAACGCCCACGCGTTGAAAGGCACCAA |
| Core Staple | AACGTGCTGAGTAGACCGTTGTACATTCT |
| Core Staple | AGACAGCATTAGCCGTACAACGCTGACGAG |
| Core Staple | CAACTGTTTCACAATTGTAATCATACGCGCGG |
| Core Staple | GAACACCCAACATATAAAACCGAGAAAGGTGA |
| Core Staple | TCTGGCCTTGAGCGAGGATCGCACCCGGAAAC |
| Core Staple | ATTAATTATCGGGAGATAATCCTGCAGATGAT |
| Core Staple | ATGAAAGTTATAGCCCCCCTCAGACATTCCAC |
| Core Staple | TAGCGACATTCATATGTATTCATTGAAACGCA |
| Core Staple | AAATATTGTAGCAAGGGCAGCACCCCATCTTT |
| Core Staple | GTCAATCAGGGATCGTACTACGAAATCTCCAA |
| Core Staple | TAAAATACAAACAAAGGAACGAGGATCAAGAG |
| Core Staple | GTAACAGTACCGCCAGGAATAGGGGATTTT |
| Core Staple | GATTATACGGATTAGAAATTTCATTTGAATTA |
| Core Staple | GGTGAGAAATGTGTAGAGGCAAGGCAAAGAAT |
| Core Staple | CGCGCAGATCATTTCACTGAATATAATGCTGT |
| Core Staple | GTTTGAAAAAAGAACGCGTTTTAATTCGAGCT |
| End Cap | CCCCGAGGACTAAAGACTTTTTCCCTGATAAATTGTCCCC |
| End Cap | CCCCTCAATAACCTGTTAACATTATGACCCTGTAATCCCC |
| End Cap | CCCCATTTATCAAAATCATAGAAGAGTCAATAGTGACCCC |
| End Cap | CCCCGTCACACGACCAGTAAATTGGCAGATTCACCACCCC |
| End Cap | AAAAGCGTAGATATTTTAAAAGTTTGAGTAACCCC |
| End Cap | CCCCTTTAGCGAACCTCCCGTAAGAACGCGAGGCGTCCCC |
| End Cap | CCCCATCAGAGCGGGAAGGGAAGACCCC |
| End Cap | CCCCGTCGAAATCCGCGTCATTACCCAAATCAACGTCCCC |
| End Cap | CCCCCCGATTGAGGGACTGGCATGATTAAGACTCCCCCC |
| End Cap | TTACTAAAAGGGAGCAATACTTCTTTGATTACCCC |
| End Cap | CCCCCAGCAGAAGATAAAACAGATTAT |
| End Cap | CCCCCACAAACAAATAAATCCTCATTAAAGCAGGTCATCTGAAAC |
| End Cap | CCCCTCTTACCAGTATAAAGCCAGACG |
| End Cap | CCCCACGCCAGCTGGCGAAAGGGGGATGTGCGCGATCGGTCGTAACC |
| End Cap | CCCCAACAAAGCTGCTCATTCAGTAAAACGAACTAACCCC |
| End Cap | CCCCTTGAGTTAAGCCCAATAGATAACCCACAAGAACCCC |
| End Cap | CCCCAAAATCCTGTTTGATGGTCCAGCTGCATTACCCC |
| End Cap | AGAGAATAAGAGCCATATTATTTATCCCAATCCCC |
| End Cap | CCCCCCAAATAAGAAACGATTTTCGGT |
| End Cap | CCCCATTGAGGAAGGTTATCTCAACAGTTGAAAGGACCCC |
| End Cap | CCCCGCGATGGCCCACTACGTGAA |
| End Cap | CCCCTCTTTCCAGACGTTAGTAAATCAGCTTGCTTTCCCC |
| End Cap | CCCCAAGCGAAAACCGTCTATCAGGCCCC |
| End Cap | CCCCAATAACCTTGCTTCTGTAATATC |
| End Cap | CCCTCAGAGCCCGTACCTATTATGACGATTGGCCTTGATATTCCCC |
| End Cap | CCCCATGAAACCGGCGACATTCAACCCC |
| End Cap | CAAATAAAATATAAAATACCGAACGAACCACCCCC |
| End Cap | CCCCATGAATCGGCCAGGTCATAGCTGTTTCCTGCCCC |
| End Cap | TCATAATCAAAATCACCGGACCCC |
| End Cap | CCGTACTCTTTCGGAATAAACAGTTAATGCCCCCCC |
| End Cap | CCCCGAAGTTTCATTCCATATAAATTTCGCAAATGGCCCC |
| End Cap | CCCCTTATAGTCAGAAGAACTAAAGTACGGTGTCTGCCCC |
| End Cap | CCCCCGAGGTGAATTTCCAACGGCTACAGAGGCTTTCCCC |
| End Cap | AGGCACCGACAATCATATGCGTTATACAAATCCCC |
| End Cap | CCCCCCGCCACCCTCAGACGATCTAAAGTTTTGTCGCCCC |
| End Cap | CCCCCCTGCCTAAGGAGGTTTAGTACCCC |
| End Cap | CTGTCGTGGGTTCCGATCCACGCTGGTTTGCCCCAGCAGGCGCCCC |
| End Cap | CCCCCAGAAATAAAGAAATTTTATTTGCACGTAAAACCCC |
| End Cap | CCCCCTATTTTTGAGAGGTAAACGTTAATATTTTGTCCCC |
| End Cap | CCCCGTAATAACATCACTTGCCTTTCCTCGTTAGACCCC |
| End Cap | CCCCACCAGAGCCACCCGTCACCACCCC |
| End Cap | CCCCTAAGAGAATATAAAGTATTTTCGAGCCAGTAACCCC |
| End Cap | CCCCCATTATCATTTTGCGGAACTTGG |
| End Cap | CCCCTGTGAAATTGTTATCCGCGGGAAGGTGCAAGGCTTGACCGT |
| End Cap | CCCCTTATTACGCAGTATGTTAGTATC |
| End Cap | CCCCAATCGGCTGTCTTTCCTTAGCAG |
| End Cap | CTGAGGTCTGAGAAATCAATATATGTGAGTGCCCC |
| End Cap | CCCCGTAGATGGGCGCATGCGGGCCTCTTCGCTATTCCCC |
| End Cap | CCCCATTCATTGAATCCCCCTCATTTACCCTGACTACCCC |
| End Cap | CCCCTAAAATTCGCATTAAATTTAGGTCACGTTGGTCCCC |
| End Cap | ATTCACTTGCGGCATGTAGAAACCAATCAATCCCC |
| End Cap | CCCCCGGAACAACATTACTGCGGAATCGTCATAAATCCCC |
| End Cap | CCCCACTTTTGCGGGAGAAGCCTCCGGAGAGGGTAGCCCC |
| Handle | CTTTTACACATTTAACGAGTACCTTTAATTGCAAAAAAAAAAAAAAA |
| Handle | GCCTGATTGAAACAAAGTCATTTTTGCGGATGAAAAAAAAAAAAAAA |
| Handle | AACTATATAAGAATAATTACCAGACGACGATAAAAAAAAAAAAAAAA |
| Handle | CGCAAGACTACCGACCAGAGGCTTTTGCAAAAAAAAAAAAAAAAAAA |
| Handle | TCCTTTTGATAAGAGCATCAAGAAAACAAAAAAAAAAAAAAAAA |
| Handle | CATCAATTCTTAATTGATTACCTGAGCAAAAGAAAAAAAAAAAAAAA |
| Handle | AAAACCAAAATAGCGGTGTGATAAATAAGGAAAAAAAAAAAAAAA |
| Handle | AAGACTTCGCCAGAGGTGACCTAAATTTAATGAAAAAAAAAAAAAAA |
| Handle | AAAGGAATTACGAGGAGAACGCGATTGTGAAAAAAAAAAAAAAA |
| Handle | GAGAATCGTCCTTGAATTAGGTTGGGTTATATAAAAAAAAAAAAAAA |
| Handle | GATGGTTTATGCGATTCAGATACATAACGCCAAAAAAAAAAAAAAAA |
| Handle | TCCAATAAATCAATAAACAATAACGGATTCAAAAAAAAAAAAAAA |
| Handle | CGTTAAATGTAAATGCCCGGAAGCAAACTCCAAAAAAAAAAAAAAAA |
| Handle | GAAGTTTTAAATATCGCGAGAAAACTTTTTCAAAAAAAAAAAAAAA |
| Handle | TTTAGACTAAAAAGATTAAGAGGAAGCCCGAAAAAAAAAAAAAAA |
| Handle | AAGATGATGCTTTGAAAGTAGTAGCATTAACAAAAAAAAAAAAAAAA |
| Fluorophore | TCAAAGCGAACCAGATGATGCAAATCCAATT(Atto 590) |